\documentclass[10pt,a4paper]{article}
\usepackage[a4paper, left=2cm, right=2cm, top=2cm]{geometry}

\usepackage[utf8]{inputenc}
\usepackage{amsmath, amsthm, amssymb, amsfonts}     
\usepackage{mathrsfs}                               
\usepackage{accents}                                
\usepackage{natbib}                                 
\usepackage{hyperref}                               
\usepackage{doi}                                    
\usepackage{authblk}                                
\usepackage{siunitx}                                
\usepackage{booktabs}                               
\usepackage{ulem}                                   
\usepackage{nomencl}                                
\makenomenclature                                   
\usepackage[linesnumbered,ruled,vlined]{algorithm2e}
\SetCommentSty{mycommfont}
\usepackage{subfigure}                              
\usepackage{soul}                                   

\usepackage{bm}                     

\newcommand{\vect}[1]{\bm{#1}} 
\newcommand{\matr}[1]{\bm{#1}} 

\usepackage{graphicx}
\usepackage{tikz}
\usepackage{pgfplots}
\usepackage[skip=0pt]{caption}
\usetikzlibrary{patterns}

\definecolor{bblue}{HTML}{4F81BD}
\definecolor{rred}{HTML}{C0504D}
\definecolor{ggreen}{HTML}{9BBB59}
\definecolor{ppurple}{HTML}{9F4C7C}

\usetikzlibrary{decorations.pathmorphing,arrows,intersections,arrows.meta,angles,quotes}
\usetikzlibrary{matrix,decorations.pathreplacing, calc, positioning}
\usepackage{multirow}
\newcommand\mydots{\ifmmode\ldots\else\makebox[0.5em][c]{.\hfil.}\fi}

\pgfplotsset{
	kurze Legende/.style={
		legend image code/.code={
			\draw[##1,mark repeat=2,line width=0.6pt]
			plot coordinates {
				(0cm,0cm)
				(0.3cm,0cm)
			};
		}
	}
}

\pgfplotsset{
	compat = newest,
	scale only axis, 
	max space between ticks = 50pt,
	ticklabel style = {font=\footnotesize},
	legend style =  {font=\footnotesize},
	grid = major,
	grid style = {dotted},
	legend columns=1, 
	xtick pos=left,
	ytick pos=left
}

\pgfplotsset{select coords between index/.style 2 args={
		x filter/.code={
			\ifnum\coordindex<#1\fi
			\ifnum\coordindex>#2\fi
		}
}}

\definecolor{color1}{HTML}{0060AD} 
\definecolor{color2}{HTML}{FF4500} 
\definecolor{color3}{HTML}{FFA500} 
\definecolor{color4}{HTML}{006400} 
\definecolor{color5}{HTML}{9400D3} 
\definecolor{color6}{HTML}{800000} 
\definecolor{color7}{HTML}{000000} 
\definecolor{color8}{HTML}{0000FF} 
\definecolor{color9}{HTML}{FF0000} 
\definecolor{mycolor_blue}{RGB}{66,124,161}
\definecolor{mycolor_grey}{RGB}{198,198,198} 

\tikzstyle{line1} = [color=color7,semithick] 
\tikzstyle{line2} = [color=color2,densely dotted,semithick]
\tikzstyle{line3} = [color=color1,densely dashed,semithick]
\tikzstyle{line4} = [color=color5,dash dot,semithick]
\tikzstyle{line5} = [color=color4,dash dot dot,semithick]
\tikzstyle{line6} = [color=color6,semithick]

\tikzstyle{mark1} = [color=color7,mark=x,mark size=2pt,mark options=solid,semithick] 
\tikzstyle{mark2} = [color=color2,mark=o,mark size=2pt,mark options=solid,semithick]
\tikzstyle{mark3} = [color=color1,mark=*,mark size=2pt,mark options=solid,semithick]
\tikzstyle{mark4} = [color=color5,mark=triangle,mark size=2pt,mark options=solid,semithick]
\tikzstyle{mark5} = [color=color4,mark=square,mark size=2pt,mark options=solid,semithick]
\tikzstyle{mark6} = [color=color7,mark=o,mark size=2pt,mark options=solid,semithick]
\tikzstyle{mark7} = [color=color7,mark=*,mark size=2pt,mark options=solid,semithick]
\tikzstyle{mark8} = [color=color7,mark=triangle,mark size=2pt,mark options=solid,semithick]

\usetikzlibrary{external}
\title{Incremental Singular Value Decomposition Based Model Order Reduction of Scale Resolving Fluid Dynamic Simulations}

\author[]{Niklas K\"uhl\thanks{mail: kuehl@hsva.de, ORCID: 0000-0002-4229-1358}}

\affil[]{Hamburg Ship Model Basin, Bramfelder Strasse 164, D-22305 Hamburg, Germany}

\date{January 20, 2024}

\begin{document}

\providetoggle{tikzExternal}
\settoggle{tikzExternal}{false}

\maketitle

\begin{abstract}
Scale-resolving flow simulations often feature several million [thousand] spatial [temporal] discrete degrees of freedom. When storing or re-using these data, e.g., to subsequently train some sort of data-based surrogate or compute consistent adjoint flow solutions, a brute-force storage approach is practically impossible. Therefore, --mandatory incremental-- Reduced Order Modeling (ROM) approaches are an attractive alternative since only a specific time horizon is effectively stored, usually aligned with the amount of fast available, e.g., Random Access Memory (RAM). This bunched flow solution is then used to enhance the already computed ROM so that the allocated memory can be released and the procedure repeats.

This paper utilizes an incremental truncated Singular Value Decomposition (itSVD) procedure to compress flow data resulting from scale-resolving flow simulations. To this end, two scenarios are considered, referring to an academic Large Eddy Simulation (LES) around a circular cylinder at $\mathrm{Re}_\mathrm{D} = 1.4 \cdot 10^5$ as well as an industrial case that employs a hybrid filtered/averaged Detached Eddy Simulation (DES) on the flow around the superstructure of a full-scale feeder ship at $\mathrm{Re}_\mathrm{L} = 5.0 \cdot 10^8$.

The paper's central focus is on an aspect of severe practical relevance: how much information of the computed scale-resolving solution should be used by the ROM, i.e., how much redundancy occurs in the resolved turbulent fluctuations that favors ROM. In the course of the tSVD employed, this goes hand in hand with the question of "how many singular values of the flow-solution-snapshot-matrix should be neglected (or considered)" -- without (a) re-running the simulation several times in a try-and-error procedure and (b) still obtain compressed results below the model and discretization error.
An adaptive strategy is utilized, which features two comparatively simple adjusting screws, for which appropriate decision support is provided. Next to a general feasibility study, reported results show the capability to obtain a fully adaptive data reduction of $\mathcal{O}(95)$ percent via a computational overhead of $\mathcal{O}(10)$ percent with a mean accuracy of reconstructed local and global flow data of $\mathcal{O}(10^{-3}-10^{-1})$ percent.

\end{abstract}

\begin{flushleft}
\end{flushleft}

\section{Introduction}


The ever-growing development of simulation software and hardware continuously increases the amount of scale-resolving flow simulations. It enables the high-fidelity determination of  \textit{global} integral quantities of interest (drag, lift, wake coefficient, etc.), both instantaneous and averaged, as well as the storage of averaged \textit{local} fields (velocity, pressure, temperature, etc.) and information derived from them, e.g., vortex criteria. However, long time series accompanied by tight Courant number restrictions are required for averaging such that (a) the storage of complete, instantaneous flow fields is --even for academic-- cases hardly possible while (b) generating a severe global footprint, especially when performed repetitively. Whenever the instantaneous flow data is to be re-used, Model Order Reduction or Reduced Order Modeling (ROM) strategies, cf. \cite{benner2015survey, mainini2015surrogate, swischuk2019projection} appear attractive and preserve dominant and, at the same time, relevant instantaneous flow information in a compressed form. Ideally, they are tailored to support conceptions of sustainable computing, e.g., consider time-to-solution and energy-to-solution, and store (or even archive) decisive high-fidelity data while neglecting redundant flow data.
The following arguments further motivates the application of ROM to scale-resolving flow simulations:
\begin{itemize}
    \item[a)] For the same starting situation (geometry, initial, and boundary conditions), different scale-resolving simulation setups (modeling, discretization, approximation) usually predict deviating integral results in the order of $\mathcal{O}(10\%)$, cf. \cite{angerbauer2020hybrid, lohner2021overnight}. Hence, the reduction/compression accuracy of instantaneous flow fields in the order of the simulation uncertainty might be sufficient for most cases. The compression effort should be increased to reach machine accuracy if needed (e.g., for the computation of consistent adjoint flow fields).
    \item[b)] Data reduction usually follows from additional computational compression effort, which seems acceptable for massively parallelized, scale-resolving flow simulations on High-Performance Computing (HPC) systems in the spirit of "computing is cheap, storing is expensive." Note that incremental, i.e., on-the-fly ROM strategies are mandatory.
\end{itemize}
A crucial part refers to the incremental, on-the-fly, or parallel-to-the-simulation construction of the ROM. Ideally, the numerical process is attached to the flow solver to minimize communication overhead and automatically detects how many modes --aligned with the number of singular values in the context of this paper-- are required to obtain a prescribed accuracy.

It is still unclear how many relevant modes are necessary to reconstruct scale-resolving flow simulations beyond academic scenarios. Certainly, there is broad agreement that the primary energy of the flow field follows the averaged velocities. Whereas turbulent fluctuations of a flow are fundamentally random, a scale-resolving discrete flow approximation usually follows a given, coded deterministic. Consequently, this determines the potential for model order reduction --e.g., via a truncated Singular Value Decomposition (tSVD), ideally below both the model and the discretization error, cf. argument a) from above-- so that finally, a reduction in storage effort in practically relevant orders of magnitude can be achieved, e.g., based on a low-rank approximation of the flow state.

The paper's applications show a reduction of $\mathcal{O}(95)$ percent with an additional effort of $\mathcal{O}(10)$ percent, providing a reconstruction accuracy of $\mathcal{O}(10^{-3}-10^{-1})$  percent, which proves the proposed incremental ROM strategy's applicability from an industrial perspective. In addition, practical tips and tricks, e.g., regarding the adaptive determination of tSVD-related truncation criteria, are provided.

The manuscript is structure as follows. Chapter 2 provides a solid overview of incremental ROM strategies in Computational Fluid Dynamics (CFD), followed by a presentation of the later employed adaptive incremental truncated Singular Value Decomposition (itSVD)  strategy introduced in Cha. 3. Subsequently, Cha. 4 presents the employed sub-grid scale models together with a description of their discrete approximation. Two case studies are performed in Cha. 5; one refers to a Large Eddy Simulation (LES) for the flow around a circular cylinder at $\mathrm{Re}_\mathrm{D} = 1.4 \cdot 10^5$, followed by the consideration of a hybrid filtered/averaged Improved Delayed Detached Eddy Simulation (IDDES) investigation on the aerodynamics of a full-scale feeder ship under generic cargo conditions at $\mathrm{Re}_\mathrm{L} = 5.0 \cdot 10^8$. Together with potential future research directions, the results are discussed in Cha. 6. Einstein's summation convention is used for Latin subscripts throughout the manuscript, and vectors as well as tensors are defined in a Cartesian coordinate system.

\section{Incremental Model Order Reduction Techniques}
\label{sec:model_order_reduction}



Incremental strategies for ROM are particularly advantageous --or even mandatory-- when the data set under investigation exceeds data processing limits (memory, input/output, etc.). Accordingly, the applications of incremental ROM extend across all scientific fields that operate with extensive data sets. Exemplary applications can be found in decision-making (i.e., recommender) systems to identify specific user/consumer behavior (\cite{sarwar2002incremental, zhou2015svd}), in the field of bio-metrics, e.g., for facial recognition (\cite{ren2010incremental, mehrotra2016incremental}), in the financial sector to support portfolio management (\cite{nakayama2003incremental, dragut2012stock}), and in general data-based learning strategies (\cite{wang2013line, zhu2017online}).

This paper employs orthogonalizing Principal Component Analysis (PCA) methods. Classical alternatives refer to Dynamic Mode Decomposition (DMD, cf. \cite{schmid2010dynamic, schmid2022dynamic}) strategies, whose application to fluid dynamics applications has been shown extensively (\cite{schmid2011applications, amor2023higher}), that, however, usually not combine (a) incremental and (b) parallelized strategies on (c) large-scale applications (\cite{reille2019incremental, calmet2020dynamic}). Applications of incremental PCA, e.g., for Proper Orthogonal Decomposition (POD) analysis of Partial Differential Equation (PDE) driven data sets, can be found, e.g., in recent work by Fareed et al. (\cite{fareed2018incremental, fareed2019note, fareed2020error}), where fundamental concepts (error analysis, temporal continuity, etc.) rather than large-scale applications are investigated. Flows with engineering complexity are usually not considered incrementally (\cite{muld2012flow}), and the focus of incremental ROM applications shifts to fluid problems with reduced complexity (e.g., 1D Burgers equation \cite{oxberry2017limited,li2021towards, li2022enhanced}, or 2D RANS studies \cite{braconnier2011towards}). Besides, serial rather than parallelized implementations are considered (\cite{fischer2023adaptive, fischer2023more}). 

More extensive, scale-resolving applications are only occasionally reported. For example, Matsumoto et al. (\cite{matsumoto2017investigation, matsumoto2019application} ) apply an itSVD strategy to the transient output of a Lattice Boltzmann Method (LBM) study for the aerodynamics of a generic automobile. However, the simulation data is (a) filtered and (b) not bunched (for multiple time steps) passed to the SVD routine. Furthermore, difficult itSVD parametrics are reported, which do not follow an adaptive strategy to determine the maximum number of singular values, which is one of the central aspects of this paper.

In the following, the incremental ROM method of this paper is described vividly; for more fundamental information, the reader should consult further literature, e.g., \cite{bunch1978updating, brand2002incremental} on mathematical or \cite{li2022enhanced, kuhl2024incremental} on algorithmic aspects.

This article's basic model reduction strategy follows from the simulation-accompanying construction of a snapshot matrix $\mathbf{Y} \in \mathbb{R}^{\mathrm{N} \times \mathrm{T}}$ so that each column corresponds to the solution vector of a computed time step. Typically, the spatial degrees of freedom far exceed the temporal ones, i.e., the snapshot matrix has significantly more rows than columns ($\mathrm{N} >> \mathrm{T}$). Note that the construction can also be transposed. Since $\mathbf{Y}$ is not square, a general eigenvalue analysis is prevented so that singular values can be considered instead. For example, scale-resolving aerodynamic studies of a ship's superstructure often involve numerical grids with approximately $\mathcal{O}(10^7)$ discrete degrees of freedom and a time horizon discretized with approximately $\mathcal{O}(10^4)$ time steps. In the case of typical hybrid averaged/filtered (Reynolds-Averaged Navier-Stokes (RANS)/Large Eddy Simulation) studies, six field variables are determined from the governing equations (three flow velocity components, one pressure variable, and two turbulence modeling parameters) so that a snapshot matrix of size $\mathbf{Y} \in \mathcal{ O} (\mathbb {R}^{6 \cdot 10^7 \times 10^4})$ has to be considered. Such a matrix can hardly be examined algebraically on a state-of-the-art workstation with reasonable effort. Even in a massively parallelized HPC scenario, storing this matrix in double precision format would require approximately $\mathcal{O}(10^1)$\footnote{$(6 \cdot 10^7 \cdot 10^4)\text{nDoF} \cdot 8 \text{Byte/nDoF} = 5 \cdot 10^{12} \text{Byte}$} terabyte of available storage capacity, which common HPC platforms usually cannot provide, especially for repetitive simulations, e.g., with varying flow conditions/parameters. In addition to spatial decomposition (e.g., via the Message Passing Interface (MPI) protocol), more sophisticated, incremental strategies are also required for the temporal evolution, cf. \cite{brand2002incremental, brand2006fast}. Such strategies only store a fraction of the actual snapshot matrix, calculate the desired SVD on the fly, and discard already processed simulation data that is no longer needed.

The SVD of a matrix ($\mathbf{Y} = \mathbf{U} \mathbf{S} \mathbf{V}^\mathrm{T}$) is exact except for machine-related rounding errors. However, since the individual singular values and the modes assigned to them have a different influence on the actual decomposition, neglecting these modes with a comparatively small contribution to the overall quality is an attractive way to store the snapshot matrix in a compressed manner with acceptable losses. Such a model reduction is usually referred to as truncated SVD (tSVD), leading to significant reductions in memory effort, especially in periodic scenarios in which states repeat, i.e. various columns of the snapshot matrix are almost identical or linearly dependent. However, it is limited in impulsive situations (e.g., traveling shock waves). In this context, a classic practical aspect refers to termination criteria that measure the number of modes considered for the tSVD, which are usually unknown a priori. These adaptive strategies relate the computed singular values and the energy of the actual snapshot matrix, which has no rigid physical meaning and is evaluated via the squared Frobenius norm, cf. \cite{grassle2018pod}. Hence, rather than specifying the exact number of singular values to be used in the reduction, a minimum number of modes (which can be set to one) combined with a minimum retained energy (typically expressed in relative terms) can be specified. Further technical aspects in this context include compliance with fundamental physical principles (e.g., positive (turbulent kinetic) energy or concentration levels), the behavior of the tSVD for state vectors of different lengths within the snapshot matrix (e.g., caused by adaptive grid refinement, cf. \cite{willcox2006unsteady}) or the construction of an all-encompassing tSVD compared to several small, individual tSVDs.

\section{Adaptive Constructed, Truncated Singular Value Decomposition}
\label{sec:incremental_svd}
\begin{figure}
    \centering
    \iftoggle{tikzExternal}{
    \input{tikz/truncated_svd.tikz}
    }{
    \includegraphics{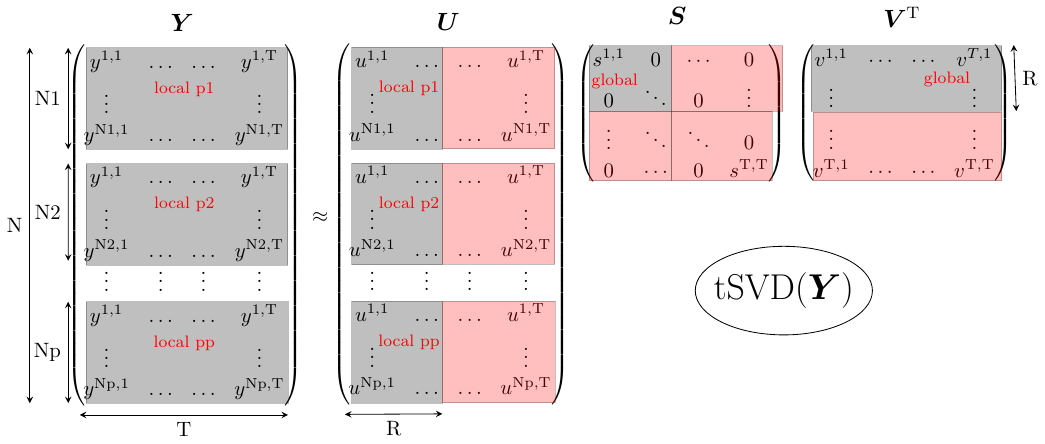}
    }
    \caption{Scetch of tSVD$\mathbf{Y}$ ($\mathrm{N} >> \mathrm{T}$) based on procedural distributed state vectors into several local sub-matrices $\mathbf{U}$ and two global matrices $\mathbf{S}$ and $\mathbf{V}$, where $\mathrm{R} < \mathrm{T}$ refers to the reduction rank.}
    \label{fig:truncated_parallel_rpts_svd}
\end{figure}

In a spatially parallelized flow simulation, the procedurally distributed state snapshot vectors (local length Np) are stitched together (global length N) as described in Sec. \ref{sec:model_order_reduction} so that the simulation history develops within the snapshot matrix $\mathbf{Y}$. An SVD-based matrix factorization approach is shown in Fig. \ref{fig:truncated_parallel_rpts_svd} ($\mathbf{Y} = \mathbf{U} \mathbf{S} \mathbf{V}^\mathrm{T}$ with $\mathbf{U} \in \mathbb{R}^{\mathrm{N} \times \mathrm{T}}$, $\mathbf{S} \in \mathbb{R}^{\mathrm{T} \times \mathrm{T}}$, and $\mathbf{V} \in \mathbb{R}^{\mathrm{T} \times \mathrm{T}}$), that also hints on a truncated approximation ($\mathbf{Y} \approx \mathbf{U} \mathbf{S} \mathbf{V}^\mathrm{T}$, with $\mathbf{U} \in \mathbb{R}^{\mathrm{N} \times \mathrm{R}}$, $\mathbf{S} \in \mathbb{R}^{\mathrm{R} \times \mathrm{R}}$, and $\mathbf{V} \in \mathbb{R}^{\mathrm{T} \times \mathrm{R}}$) by deliberately neglecting the rearward, less relevant $\mathrm{T}-\mathrm{R}$ singular values within the centered matrix $\mathbf{S}$.
Due to the spatially parallel but temporally serial simulation, $\mathbf{S}$ as well as $\mathbf{V}$ coincide on all processors --required, e.g., by domain decomposition strategies-- which requires a careful implementation with special consideration of parallel communication, see \cite{kuhl2024incremental}. The reduction rank R serves as a reducing element, is usually not known a priori and has to be adjusted adaptively. Since the first singular values usually feature extreme numerical dominance, careless handling of the matrix energy can quickly become an issue of machine accuracy, especially since the singular values are often considered squared. Therefore, a modified, heuristic approach has proven helpful, which deliberately ignores the first $o$ singular values during the adaptive adjustment and works exclusively on the remaining energy, cf.
\begin{align}
    \eta^\mathrm{o} = \frac{\sum_\mathrm{i=o}^\mathrm{R} s_\mathrm{i}^2}{E - \sum_\mathrm{i=1}^\mathrm{o-1} s_\mathrm{i}^2}
    \qquad \text{with} \qquad 
    E = \sum_\mathrm{i=1}^\mathrm{T} ||\mathbf{y}_\mathrm{i}||^2
    \qquad \qquad \qquad \overset{o = 1}{\to} \qquad \qquad \qquad 
    \eta = \frac{\sum_\mathrm{i=1}^\mathrm{R} s_\mathrm{i}^2}{E}
    \label{equ:adaptive_energy} \, .
\end{align}
Therein, $s_\mathrm{i}$, $o$ and $\mathrm{R}$ refer to the singular values, and minimum as well as variable (maximum) reduction rank, respectively. The retained energy refers to $\eta^\mathrm{o} = [0,1]$ and recovers classic approximation measures (\cite{gubisch2017proper, grassle2018pod}) in the limit of $o=1$. Compromising the exact energy content is not needed here since the exact matrix energy can be determined in parallel with the time integration or snapshot matrix evolution by incrementally summing up all instantaneous energies.

A practical implementation, therefore, requires the specification of the minimum reduction rank $o$ and the remaining energy content $\eta^\mathrm{o}$. The reduction rank $R \geq o$ is increased until the matrix energy to be maintained is reached. Relevant tips and tricks for an appropriate choice of $o$ and $\eta^\mathrm{o}$ are discussed in this paper's application and conclusion part.

\section{Scale Resolving Flow Simulation Models \& Numerical Approximation Strategy}
Two prominent scale-resolving eddy-viscosity approaches are considered, i.e., (a) a classical LES approach and (b) a hybrid RANS/LES method.
\begin{enumerate}
    \item[a)] As a representative of classical LES approaches, the Wall-Adapting Local Eddy Viscosity (WALE) model of \cite{ducros1998wall, nicoud1999subgrid} is employed in its static form, i.e., with a constant sub-grid length scale model parameter that is set to $C^\mathrm{w} = 0.325$. Further details are provided in App. \ref{subsec:large_eddy_simulation}.
    \item[b)] Secondly, a hybrid filtered/averaged DES (\cite{spalart2009detached}) approach is used for turbulence modeling. The Improved Delayed Detached Eddy Simulation (IDDES) model is applied in line with a variant of \cite{gritskevich2012development} that builds on the popular $k$-$\omega$ Boussinesq viscosity approach of Menter's Shear Stress Transport (SST) model, cf. \cite{menter1994two, menter2003ten}. Default values are used for all constants in the model. Further details are provided in App. \ref{subsec:detached_eddy_simulation} and validation studies can be found in \cite{schrader2018aerodynamics, angerbauer2020hybrid, pache2022data, loft2023two}.
\end{enumerate}
The numerical procedure employs the Finite Volume Method (FVM) FreSCo$^\text{+}$ on a parallel CPU machine with distributed memory, cf. \cite{stuck2012adjoint, kroger2016numerical, manzke2018development, schubert2019analysis, kuhl2021phd}. The approach is based on a second-order approximation associated with arbitrary unstructured grids, and the parallelization follows a domain decomposition approach, as described in \cite{yakubov2013hybrid, yakubov2015experience}. The algorithm is based on a cell-centered and co-located arrangement of variables. Temporal derivatives are approximated by an implicit method with three time levels and spatial integrals by the midpoint rule. Second-order central differences approximate diffusive fluxes, and higher-order (bounded) upwind-biased methods are generally employed to reconstruct convective fluxes. A modified pressure correction algorithm accounts for the odd-even decoupling of pressure and velocity components, cf. \cite{kuhl2022discrete}.

Values within the local state vectors from Sec. \ref{sec:model_order_reduction} are passed to the solver's itSVD subroutine at the end of each time step. Related pseudo code lines are provided in Algs. \ref{alg:tSVD_construction}-\ref{alg:tSVD_evaluation}, where Alg. \ref{alg:tSVD_adaption} mimics the adaptation process, cf. Eqn. \ref{equ:adaptive_energy}. After bunching a certain amount of timesteps ($b \approx 500$ in Alg. \ref{alg:tSVD_construction}), they are appended to the previous tSVD, cf. \cite{kuhl2024incremental}. Once the itSVD construction has been performed, the evaluation part enrolls the tSVD data in time, allowing for a quality check of the reduced data's accuracy. 

\begin{algorithm}[!ht]
\SetKwInput{KwInput}{Input}
\SetKwInput{KwOutput}{Output}
\SetKwInput{KwOutputOpt}{Optional Output}
\DontPrintSemicolon

\KwInput{$b, o, \eta^o\in \mathbb{N}$\tcp*{Bunch size, minimum truncation rank, retained energy}}
\KwOutput{$\matr{U} \in \mathbb{R}^\mathrm{Y \times R}$, $\matr{S} \in \mathbb{R}^\mathrm{R,R}$, $\matr{V} \in \mathbb{R}^\mathrm{T, R}$\tcp*{itSVD matrices}}
\KwOutputOpt{$\matr{F} \in \mathbb{R}^{\mathrm{T}, h}$\tcp*{h quantities of interest (e.g. drag/lift coefficient) over time}}

\SetKwFunction{FMain}{construct{\_}SVD}
\SetKwFunction{FUpdate}{itSVD{\_}update}
 
\SetKwProg{Fn}{Module}{:}{\KwRet}
\Fn{\FMain{}}{
    $k = 0$\tcp*{Initialize the bunch matrix counter}
    \For{t=1,T} 
    {
    	    compute state vectors $\vect{\varphi}(:)$\tcp*{Compute actual state vector, i.e., solve LES/DES system}
            determine $\matr{F}(t,:)$\tcp*{Determine quantities of interest}
    	    $E \leftarrow (\vect{\varphi}(:)^\mathrm{T}\vect{\varphi}(:))$\tcp*{Update the global matrix energy, cf. Eqn. \ref{equ:adaptive_energy}}
            $k++$\tcp*{Increase the bunch matrix counter}
            $\matr{B}(:,k) \leftarrow \vect{\varphi}(:)$ \tcp*{Update the bunch matrix by state $\varphi$}
            \If{(k=b) \textbf{or} (t=T)}
            {
                $[\matr{U}^\prime, \matr{S}^\prime, {\matr{V}^\mathrm{T}}^\prime] =$itSVD{\_}in($\matr{U}, \matr{S}, \matr{V}^\mathrm{T}, \matr{B}$) \tcp*{Intermediate tSVD, cf. \cite{kuhl2024incremental}}
                R = adaptive{\_}truncation($\matr{S}^\prime$)\tcp*{Adapt the truncation rank, cf. Eqn. \ref{equ:adaptive_energy}}
                $[\matr{U}, \matr{S}, \matr{V}^\mathrm{T}]$ = itSVD{\_}up $[\matr{U}^\prime, \matr{S}^\prime, {\matr{V}^\mathrm{T}}^\prime](1:R,1:R)$ \tcp*{Update tSVD, cf. \cite{kuhl2024incremental}}
                $\matr{B}(:,:) = 0$, $k = 0$ \tcp*{Clear the bunch matrix and its counter}
            }
    }
}
\caption{Pseudo code example for the construction process of an incremental truncated Singular Value Decomposition during the time integration process. More details are provided in \cite{kuhl2024incremental}.}
\label{alg:tSVD_construction}
\end{algorithm}

\begin{algorithm}[!ht]
\SetKwInput{KwInput}{Input}
\SetKwInput{KwOutput}{Output}
\DontPrintSemicolon


\SetKwFunction{FMain}{adaptive{\_}truncation}
 
\SetKwProg{Fn}{Function}{:}{\KwRet}
\Fn{\FMain{$\matr{S}$}}{
    \eIf{size($\matr{S}$,1) $\leq$ o}
    {
        $R = \mathrm{size}(\matr{S},1)$\tcp*{Use all singular values}
    }{
        \For{R=o,size($\matr{S}$,1)}
        {
            \If{$ [ {\matr{S}}(o:R,o:R)\matr{S}(o:R,o:R) ] / [E - {\matr{S}}^\mathrm{T}(1:o-1,1:o-1)\matr{S}(1:o-1,1:o-1) ] \geq \eta^\mathrm{o}$}{exit\tcp*{Increase $R$ and leave when the energetic bound is met, cf. Eqn. \ref{equ:adaptive_energy}}}
        }
    }
    \KwRet $R$
}
\caption{Pseudo code example for the adaptive computation of the truncation rank $R$, cf. Eqn. \ref{equ:adaptive_energy}.}
\label{alg:tSVD_adaption}
\end{algorithm}

\begin{algorithm}[!ht]
\SetKwInput{KwInput}{Input}
\SetKwInput{KwOutputOpt}{Optional Output}
\DontPrintSemicolon

\KwInput{$\matr{U} \in \mathbb{R}^\mathrm{Y \times R}$, $\matr{S} \in \mathbb{R}^\mathrm{R,R}$, $\matr{V} \in \mathbb{R}^\mathrm{T, R}$\tcp*{itSVD matrices}}
\KwOutputOpt{$\tilde{\matr{F}} \in \mathbb{R}^{\mathrm{T}, h}$\tcp*{h quantities of interest (e.g. drag/lift coefficient) over time}}

\SetKwFunction{FMain}{evaluate{\_}SVD}
 
\SetKwProg{Fn}{Function}{:}{\KwRet}
\Fn{\FMain{}}{
    \For{t=1,T} 
    {
        $\tilde{\vect{\varphi}}(:) \leftarrow \matr{U}(:,1:R) \matr{S}(1:R,1:R) \matr{V}(t,1:R) $ \tcp*{Evaluate the tSVD as an approximation of state $\tilde{\varphi}$}
        determine $\tilde{\matr{F}}(t,:)$\tcp*{Determine (approximated) quantities of interest}
    }
}
\caption{Pseudo code example for the (temporal) evaluation of a truncated Singular Value Decomposition.}
\label{alg:tSVD_evaluation}
\end{algorithm}

\section{Case Studies}
The outlined itSVD procedure is applied to two resolved flow studies in this section that refer to an academic flow around a circular cylinder and an industrial scenario considering the aerodynamics of a feeder vessel with a generic container distribution. For varying adaptivity parameters in Eqn. \eqref{equ:adaptive_energy}, instantaneous integral data as drag ($c^\mathrm{d}$) and lift ($c^\mathrm{l}$) coefficient are examined, viz.
\begin{align}
c^\mathrm{d} = \frac{2 f^{\parallel}}{\rho A V^2},
\qquad
c^\mathrm{l} = \frac{2 f^{\perp}}{\rho A V^2},
\qquad \qquad \mathrm{with} \qquad \qquad
f_\mathrm{i} = \int_{\Gamma^\mathrm{w}} \left[ p^\mathrm{eff} \delta_{ik} - \mu^\mathrm{eff} \left( \frac{\partial v_\mathrm{i}}{\partial x_\mathrm{k}} + \frac{\partial v_\mathrm{k}}{\partial x_\mathrm{i}} \right) \right] n_\mathrm{k} \mathrm{d} \Gamma \, , \label{eqn:drag_lift_coefficient}
\end{align}
where $v_i$ and $p^\mathrm{eff}$ refer to the fluid's velocity and (effective) pressure field, respectively, cf. App. \ref{app:governing_equations} for more details on the underlying flow model. Above, $f^{\parallel}$ and $f^{\perp}$ refer to the bulk flow longitudinal and lateral force components that follow from a projection of the flow-induced force $f_i$ along all wall boundaries $\Gamma^\mathrm{w}$. In addition, instantaneous local flow data extracted at specific probe locations are considered. Quality metrics refer to the error of reduced ($\tilde{\varphi}$) local/integral w.r.t. simulated data ($\varphi$), i.e., aim at a relative error that reads $e^\mathrm{\varphi} = (\tilde{\varphi} - \varphi)/\varphi \cdot 100\%$. Furthermore, storing related aspects (required memory) and simulation (itSVD overhead) are considered.


\subsection{Circular Cylinder}
\label{subsec:cylinder}
\begin{figure}[!ht]
\centering
\subfigure[]{
\iftoggle{tikzExternal}{
\input{./tikz/cylinder/cylinder_sketch.tikz}
}{
\includegraphics{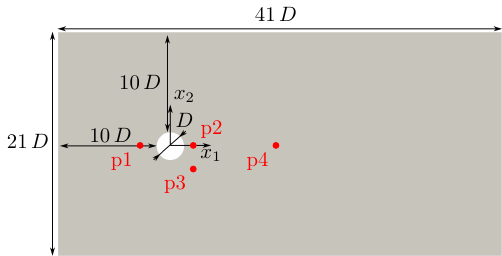}
}
}
\subfigure[]{
\includegraphics[width=0.425\textwidth]{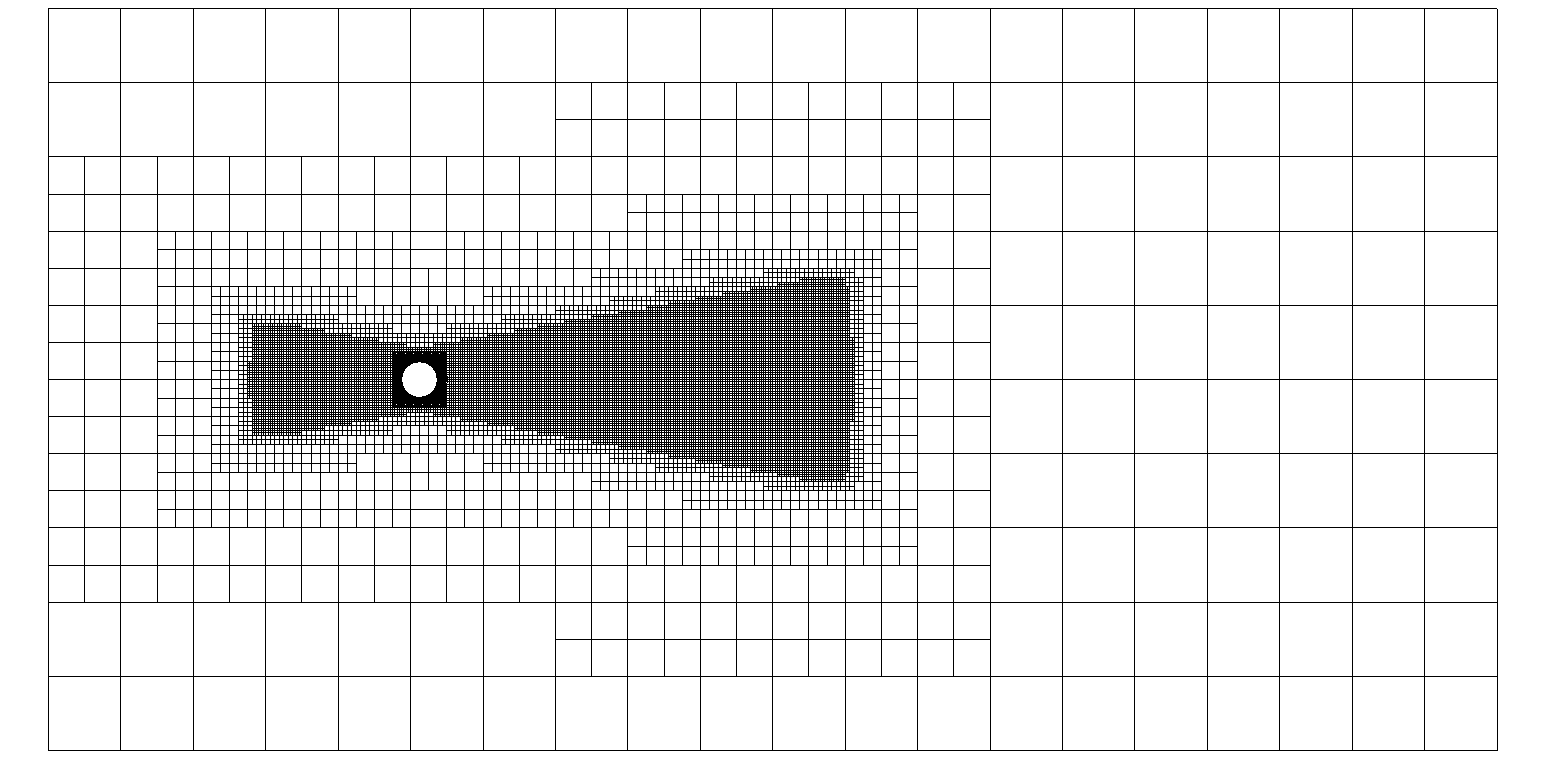}
}
\caption{Cylinder flow ($\mathrm{Re}_\mathrm{D} = 1.4 \cdot 10^5$): (a) Schematic 2D drawing and (b) numerical grid in the center plane $x_3/D = 1$.}
\label{fig:cylinder_sketch}
\end{figure}

The paper's pure LES investigation considers a single-phase (kinematic viscosity $\nu$) flow around a submerged circular cylinder of diameter $D$ at $\mathrm{Re}_\mathrm{D} = V D / \nu = \SI{140000}{}$ based on the bulk inlet velocity $V$. The test case has been intensively studied both experimentally (\cite{achenbach1968distribution, son1969velocity, cantwell1983experimental}) and numerically (\cite{breuer2000challenging, kim2013characteristics, pereira2019simulation}). The considered domain is sketched in Fig. \ref{fig:cylinder_sketch} (a) and features a length/height/width of [$41 \, D$, $21 \, D$, $2 \, D$] where the inlet and bottom boundaries are ten diameters away from the cylinder's axis. A homogeneous unidirectional (horizontal) bulk flow $v_\mathrm{i} = V \delta_{i1}$ is imposed along the inlet, and a constant pressure $P$ is prescribed along with the downstream located outlet. All remaining boundaries are assigned to symmetry planes. Initial values are aligned with the inlet/outlet condition and follow a constant pressure field and a homogeneous, horizontal flow.
A slice along the cylinder's center plane at $x_\mathrm{3}/D = 1$ of the utilized unstructured numerical grid is displayed in Fig. \ref{fig:cylinder_sketch} (b). The numerical grid consists of approximately $\SI{1 600 000}{}$ control volumes where the cylinder shape is discretized with $90 \times 80$ elements in circumferential and span-wise directions. The non-dimensional wall-normal distance of the first grid layer reaches down to $y^+ \approx \SI{0.5}{}$, and the grid is refined inside a box of size [$1.2 \, D \times 1.2 \, D \times 2 \, D$] with a spacing of $\Delta x_\mathrm{1} / D = 0.03, \Delta x_\mathrm{2} / D = 0.03, \Delta x_\mathrm{3} / D = 0.03$. An additional downstream located refinement sector with an opening angle of $20^\circ$ and a length and height of $15 \, D$ as well as $2 \, D$, respectively, accounts for the expected vortex street with $\Delta x_\mathrm{1} / D = 0.09, \Delta x_\mathrm{2} / D = 0.09, \Delta x_\mathrm{3} / D = 0.09$.
All convective and diffusive fluxes are approximated via a Central Difference Scheme (CDS) and the second-order Implicit Three Time Level (ITTL) integration utilizes a constant temporal increment of $\Delta t / (D / V) = 3 / 1000$, i.e., one advective passage is discretized with 333 time steps, ensuring a local cell Courant number well below one, i.e., $\mathrm{Co} = \mathcal{O}(0.5)$.
A time horizon of $\SI{50 000}{}$ time steps is simulated to obtain statistically converged flow results, corresponding to approximately 150 equivalent flows. The considered model order reduction does not address the initial transient part and becomes active after 30 advective passings, i.e., $t^\mathrm{s}/(D/V_\mathrm{1}) = 30$ and $t^\mathrm{e}/(D/V_\mathrm{1}) = 150$ or $T = \SI{40 000}{}$ in Eqn. \eqref{equ:adaptive_energy}.

In addition to the integral evaluation criteria from Eqn. \eqref{eqn:drag_lift_coefficient}, the model reduction of the cylinder flow is evaluated using an additional integral cost function, which addresses the mean deviation of the instantaneous flow velocity $v_\mathrm{i}$ from a target companion $v_\mathrm{i}^\mathrm{tar} = \delta_{i1}$, viz.
\begin{align}
c^\mathrm{v} = \frac{1}{V^2 A D} \int_{\Omega^\mathrm{O}} \frac{1}{2} \left[ v_\mathrm{i} - v_\mathrm{i}^\mathrm{tar} \right]^2 \mathrm{d} \Omega \, .
\end{align}
The deviation is evaluated inside a boxxed region of $\Omega^\mathrm{O} = [1.5 D, -D, 0] \times [4D, D, 2D]$ and non-dimensionalized using the projected reference area $A = 2 D^2$ times the cylinder diameter $D$ and the bulk flow velocity $V$.

Twelve different numerical experiments are carried out, which, following Eqn. \eqref{equ:adaptive_energy}, vary both the minimum ROM rank $o = [100, 200, 400, 800]$ and, based on this, the minimum energy to be maintained $\eta^\mathrm{o} \cdot 100 \% = [10, 50, 90]$, resulting in 12 different, adaptively determined tSVD ranks $R$. The quality of these model reductions is measured using integral (drag, lift, and velocity deviation coefficient) and local (pressure, velocity probe) data. Four probes are positioned along the domain mid-plane $x_\mathrm{3}/D = 1$ as shown in Fig. \ref{fig:cylinder_sketch}, with one immediately upstream ($x_\mathrm{1}/D = -1.0$, $x_\mathrm{2}/D = 0.0$) as well as downstream of the cylinder in its wake ($x_\mathrm{1}/D = 1.0$, $x_\mathrm{2}/D = 0.0$), one below ($x_\mathrm{1}/D = 1.0$, $x_\mathrm{2}/D = -1.0$), and one significantly further downstream ($x_\mathrm{1}/D = 5.0$, $x_\mathrm{2}/D = 0.0$) of the cylinder. Two separate tSVD's are generated for the scalar pressure field $p$ and the vector-valued velocity field $v_\mathrm{i}$, cf. \cite{kuhl2023archiving}.

\begin{figure}[!ht]
\centering
\iftoggle{tikzExternal}{
\input{./tikz/cylinder/cylinder_singular_value_adaption.tikz}
}{
\includegraphics{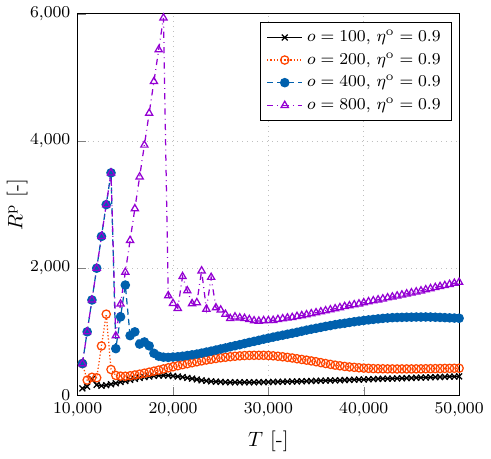}
\includegraphics{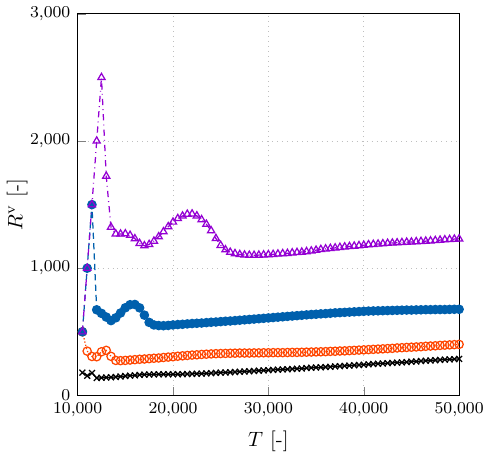}
}
\caption{Cylinder flow ($\mathrm{Re}_\mathrm{D} = 1.4 \cdot 10^5$): Number of considered singular values of the pressure (left, $R^\mathrm{o, p}$) and velocity (right, $R^\mathrm{o, v}$) field over the simulated time steps for minimal construction ranks $o=[100, 200, 400, 800]$ with a retained energy of $\eta^\mathrm{o} = 0.9$ each.}
\label{fig:cylinder_singular_value_adaption}
\end{figure}

\begin{figure}[!ht]
\centering
\iftoggle{tikzExternal}{
\input{./tikz/cylinder/cylinder_singular_value_energy.tikz}
}{
\includegraphics{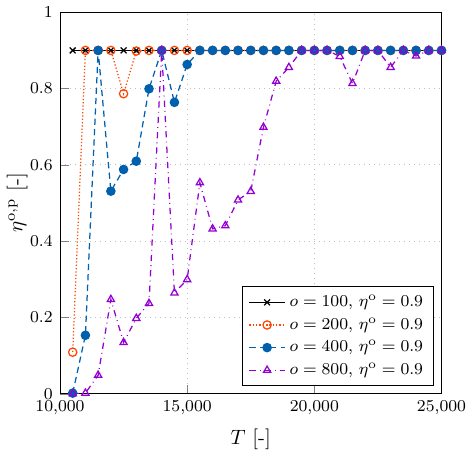}
\includegraphics{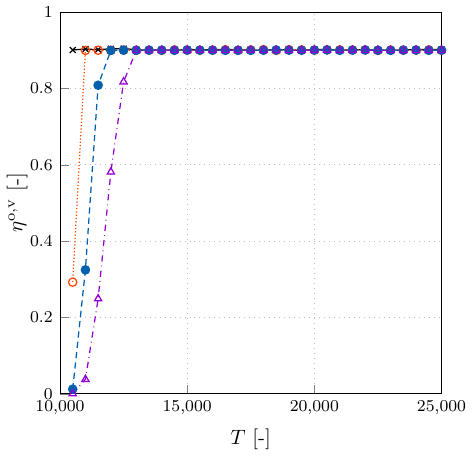}
}
\caption{Cylinder flow ($\mathrm{Re}_\mathrm{D} = 1.4 \cdot 10^5$): Retained energy of the pressure (left, $\eta^\mathrm{o, p}$) and velocity (right, $\eta^\mathrm{o, v}$) field over the simulated time steps for minimal construction ranks $o=[100, 200, 400, 800]$ with a retained energy of $\eta^\mathrm{o} = 0.9$ each.}
\label{fig:cylinder_singular_value_energy}
\end{figure}

The number of singular values $\mathrm{R}^\mathrm{\varphi}$, cf. Eqn. \eqref{equ:adaptive_energy}, determined adaptively, is given in Tab. \ref{tab:cylinder_svd_results} for all 12 scenarios, together with the actually retained energy $\tilde{\eta}^\mathrm{o, \varphi}$ for both the vector-valued velocity ($\varphi \to v$) and the scalar pressure ($\varphi \to p$) field. Additionally, the construction time required $t^\mathrm{svd, \varphi}$ (normalized with the total simulation time of the flow solver $t^\mathrm{cfd}$) as well as the required storage $d^\mathrm{svd, \varphi}$ (normalized with the capacity of a full-storage storage approach $d^\mathrm{full, \varphi}$) are provided.
\begin{table}[!ht]
\caption{Cylinder flow ($\mathrm{Re}_\mathrm{D} = 1.4 \cdot 10^5$): Number of adaptively computed singular values $\mathrm{R}^\mathrm{\varphi}$ together with the retained energy $\tilde{\eta}^\mathrm{o, \varphi}$, cf. Eqn. \eqref{equ:adaptive_energy}, for both, the vector-valued velocity ($\varphi \to v$) and the scalar pressure ($\varphi \to p$) field. In addition, the relative simulation effort $t^\mathrm{svd, \varphi}$ (normalized with the simulation time of the flow solver $t^\mathrm{cfd}$) and storage overhead $d^\mathrm{svd, \varphi}$ (normalized with the required storage capacity of a full storage approach $d^\mathrm{full, \varphi}$) are provided.}
\begin{center}
\begin{tabular}{c|ccc|ccc|ccc|ccc}
\toprule
$o$ & \multicolumn{3}{c|}{100} & \multicolumn{3}{c|}{200} & \multicolumn{3}{c|}{400} & \multicolumn{3}{c}{800} \\ \midrule 
$\eta^\mathrm{o} \cdot 100\%$ & 10.00 & 50.00 & 90.00 & 10.00 & 50.00 & 90.00 & 10.00 & 50.00 & 90.00 & 10.00 & 50.00 & 90.00 \\ \midrule \midrule
$\mathrm{R}^\mathrm{v}$ & 107 & 153 & 287 & 208 & 259 & 401 & 410 & 469 & 677 & 817 & 919 & 1231 \\ \midrule 
$\tilde{\eta}^\mathrm{o, v} \cdot 100\%$ & 10.03 & 50.43 & 90.09 & 10.12 & 50.36 & 90.03 & 10.80 & 50.34 & 90.01 & 10.14 & 50.11 & 90.01 \\ \midrule 
$t^\mathrm{svd, v}/t^\mathrm{cfd} \cdot 100\%$ & 2.45 & 2.76 & 2.83& 6.42 & 6.69 & 8.04 & 9.33 & 10.14 & 12.58 & 16.43 & 16.69 & 21.08 \\ \midrule 
$d^\mathrm{svd, v}/d^\mathrm{full, v} \cdot 100\%$ & 0.61 & 0.66 & 0.81 & 0.72 & 0.77 & 0.90 & 0.93 & 0.99 & 1.21 & 1.33 & 1.33 & 1.52 \\ \midrule \midrule
$\mathrm{R}^\mathrm{p}$ & 108 & 157 & 298 & 209 & 262 & 427 & 413 & 505 & 1213 & 859 & 1180 & 1779 \\ \midrule 
$\tilde{\eta}^\mathrm{o, p} \cdot 100\%$ & 10.10 & 50.20 & 90.01 & 10.50 & 50.13 & 90.01 & 10.52 & 50.24 & 90.00 & 10.03 & 50.05 & 90.00 \\ \midrule 
$t^\mathrm{svd, p}/t^\mathrm{cfd} \cdot 100\%$ & 1.32 & 1.67 & 2.06 & 3.71 & 4.30 & 6.78 & 6.16 & 7.31 & 14.58 & 11.72 & 13.67 & 19.34 \\ \midrule 
$d^\mathrm{svd, v}/d^\mathrm{full, p} \cdot 100\%$ & 0.20 & 0.22 & 0.27 & 0.24 & 0.26 & 0.30 & 0.31 & 0.33 & 0.40 & 0.44 & 0.48 & 0.51 \\ \midrule 
\bottomrule
\end{tabular}
\end{center}
\label{tab:cylinder_svd_results}
\end{table}
The number of singular values reaches their minimum [maximum] for the lowest [highest] minimum construction ranks ($o=100$) [($o=800$)], as well as the smallest ($\eta^\mathrm{o} = 0.1$) [largest ($\eta^\mathrm{o} = 0.9$)] energy to be retained. This can also be seen in Fig. \ref{fig:cylinder_singular_value_adaption}, which provides the adaptively determined number of singular values of the pressure (left, $R^\mathrm{p}$) and the velocity field (right, $R^\mathrm{v}$) over the number of the itSVD-relevant time steps for different minimum construction ranks $o=[100, 200, 400, 800]$ and a retained energy of $\eta^\mathrm{o} = 0.9$. Especially for high $o$-values, a constant increase in singular values arises at the beginning of the itSVD construction, which is attributed to the prescribed energy limit not being maintained at this simulation time. The retained energy is plotted in Fig. \ref{fig:cylinder_singular_value_energy} for the pressure (left, $\eta^\mathrm{o, p}$) and velocity (right, $\eta^\mathrm{o, v}$) fields over the time steps. Approximately 15-25 [5] percent of the simulation time is required to reach the pressure [velocity] field's energy limit, which underlines the almost constant increase of singular values in this time interval.
The remaining ten scenarios yield values between the two extreme cases, where compared to higher energy constraints, increased lower rank limits tend to result in an increased number of singular values. The simulation and storage overheads match these observations. They scale with the number of singular values considered, with a time overhead between 2\% for a small number of singular values and about 10-20\% for the higher reduction cases. The storage overhead is in the range of 0.1-1.5\%, although compression effects due to the employed H5 library should be considered, which may differ from tSVD-only-storage to the full-storage case. Data for the latter is known from another publication (currently under review) and amounts to approx. 0.7 terabyte per field quantity and \SI{40000}{} time steps. Hence, storing the entire velocity field requires about 2.1 terabytes of disc space. Note that simulation and memory overheads depend on hardware (inter-processor communication, etc.) and software (algorithms, etc.) aspects, so the provided data are only indicative.

\begin{figure}[!ht]
\centering
\iftoggle{tikzExternal}{
\input{./tikz/cylinder/cylinder_integral_error.tikz}
}{
\includegraphics{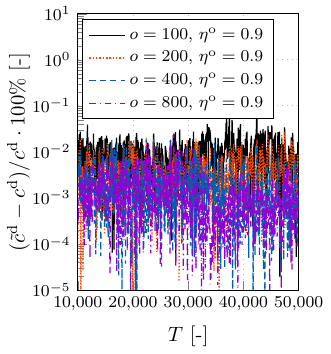}
\includegraphics{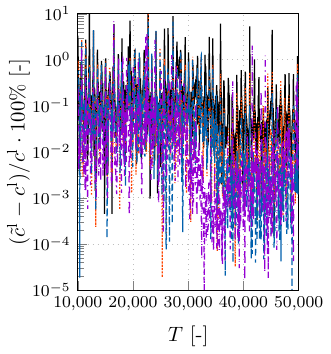}
\includegraphics{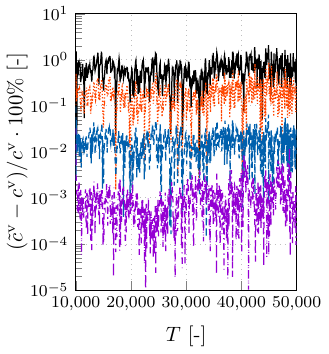}
}
\caption{Cylinder flow ($\mathrm{Re}_\mathrm{D} = 1.4 \cdot 10^5$): Instantaneous erros $(\tilde{c}^{(\cdot)} - c^{(\cdot)})/c^{(\cdot)} \cdot 100\%$ of reconstructed drag (left), lift (center), and velocity deviation (right) coefficient for minimal construction ranks $o=[100, 200, 400, 800]$ with a retained energy of $\eta^\mathrm{o} = 0.9$ each.}
\label{fig:cylinder_integral_errors}
\end{figure}

\begin{figure}[!ht]
\centering
\iftoggle{tikzExternal}{
\input{./tikz/cylinder/cylinder_integral_error_mean.tikz}
}{
\includegraphics{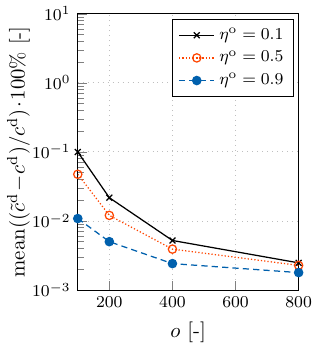}
\includegraphics{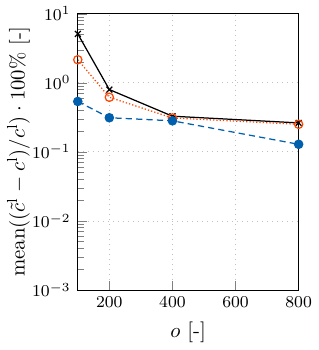}
\includegraphics{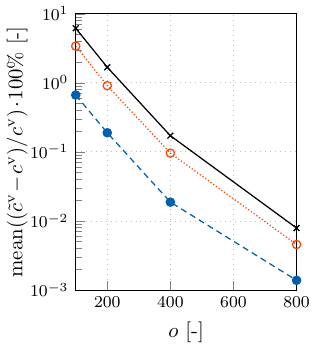}
}
\caption{Cylinder flow ($\mathrm{Re}_\mathrm{D} = 1.4 \cdot 10^5$): Time-averaged reconstruction errors ($\text{mean}((\tilde{c}^{(\cdot)} - c^{(\cdot)})/c^{(\cdot)} \cdot 100\%)$) of compressed drag (left), lift (center), and velocity deviation (right) coefficient over the minimal construction rank for different retained energy levels $\eta^\mathrm{o}=[0.1,0.5,0.9]$.}
\label{fig:cylinder_integral_error_mean}
\end{figure}

Results of several reduced order models are considered in the following based on a re-evaluation of the compressed flow fields while comparing against the simulated reference data. First, the instantaneous error $(\tilde{c}^{(\cdot)} - c^{(\cdot)})/c^{(\cdot)} \cdot 100\%$) of the reconstructed drag ($\tilde{c}^\mathrm{d}$, left), lift ($\tilde{c}^\mathrm{l}$, center), and velocity deviation ($\tilde{c}^\mathrm{v}$, right) coefficient is shown Fig. \ref{fig:cylinder_integral_errors} for the maximum retained energy $\eta^\mathrm{o} = 0.9$ and four varying minimum construction ranks $o=[100, 200, 400, 800]$ as black-solid, orange-dotted, blue-dashed, and purple-dash-dotted line, respectively. In all cases, the errors decrease for higher minimum ranks (and therefore more singular values used; cf. Tab. \ref{tab:cylinder_svd_results}) and are in the range of $\mathcal{O}(10^{-3})\%$ for $c^\mathrm{d}$, $c^\mathrm{v}$, and $\mathcal{O}(10^{-2})\%$ for $c^\mathrm{l}$. The increased error in the lift reconstruction follows from a normalization being susceptible to singularities ($c^\mathrm{l} \approx 0$), further underlined by the minor difference between the four error curves in the center figure. Lower minimum construction ranks tend to increase the reconstruction errors, with the differences in the respective reconstruction quality for the inverse velocity functional appearing particularly extreme. This also becomes clear in Fig. \ref{fig:cylinder_integral_error_mean}, in which the time-averaged reconstruction errors are plotted over the minimum construction rank for all considered cases, i.e., with varied retained energies ($\eta^\mathrm{o, \varphi} = 0.1$:black-solid, $\eta^\mathrm{o, \varphi} = 0.5$:orange-dotted, $\eta^\mathrm{o, \varphi} = 0.9$:blue-dashed). The averaged errors decrease for higher $o$-values and increased retained energies, with the slope being most extreme in the case of the inverse velocity functional. Note that the habitat of the latter functional covers an increased spatial domain compared to the force functional. Regarding the force coefficients, doubling the minimum design rank of $o=400$ results in minimal improvements in the respective averaged error. To conclude, in between, a good candidate for the minimum reduction rank follows from the number of advective time steps (333 here) to compress integral quantities adequately.

\begin{figure}[!ht]
\centering
\iftoggle{tikzExternal}{
\input{./tikz/cylinder/cylinder_prope2_error.tikz}
}{
\includegraphics{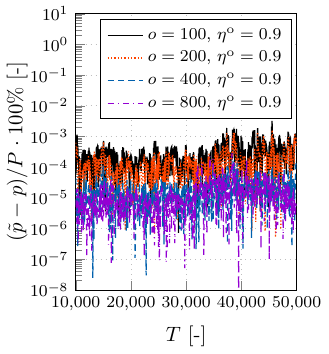}
\includegraphics{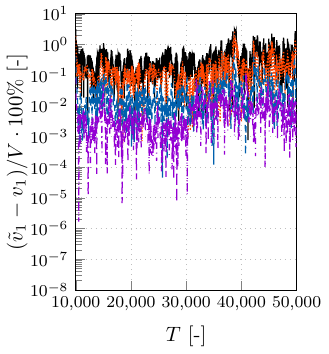}
\includegraphics{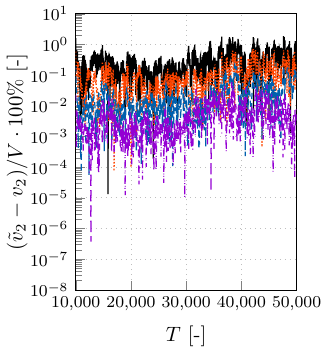}
}
\caption{Cylinder flow ($\mathrm{Re}_\mathrm{D} = 1.4 \cdot 10^5$): Instantaneous erros ($(\tilde{\varphi} - \varphi)/\varphi^\mathrm{ref} \cdot 100\%$) of reconstructed pressure (left), horizontal (center), and vertical (right) velocity for minimal construction ranks $o=[100, 200, 400, 800]$ with a retained energy of $\eta^\mathrm{o} = 0.9$ at probe 2 ($x_\mathrm{1}/D = 1.0$, $x_\mathrm{2}/D = 0.0$, $x_\mathrm{3}/D = 1.0$).}
\label{fig:cylinder_local_errors_probe_2}
\end{figure}

\begin{figure}[!ht]
\centering
\iftoggle{tikzExternal}{
\input{./tikz/cylinder/cylinder_prope2_error_mean.tikz}
}{
\includegraphics{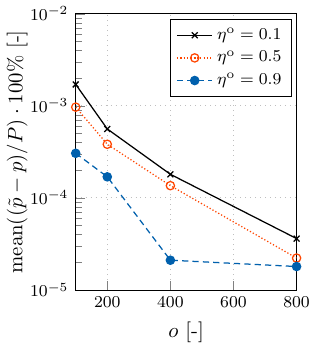}
\includegraphics{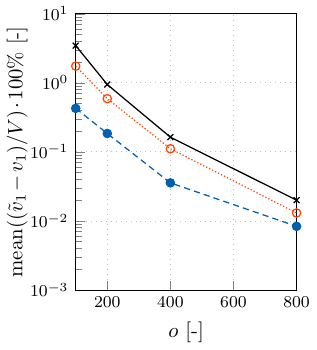}
\includegraphics{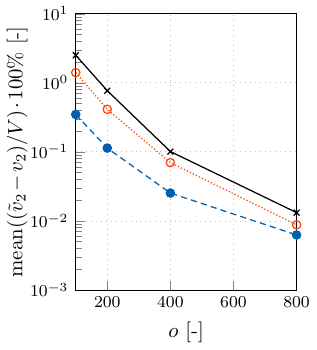}
}
\caption{Cylinder flow ($\mathrm{Re}_\mathrm{D} = 1.4 \cdot 10^5$): Time-averaged reconstruction errors ($\text{mean}((\tilde{\varphi} - \varphi)/\varphi^\mathrm{ref} \cdot 100\%)$) of compressed pressure (left), horizontal (center), and vertical (right) velocity over the minimal construction rank for different retained energy levels $\eta^\mathrm{o}=[0.1,0.5,0.9]$ at probe 2 ($x_\mathrm{1}/D = 1.0$, $x_\mathrm{2}/D = 0.0$, $x_\mathrm{3}/D = 1.0$).}
\label{fig:cylinder_prope2_error_mean}
\end{figure}

For the same ROM scenarios with $\eta^\mathrm{o} = 0.9$ and $o=[100, 200, 400, 800]$, Fig. \ref{fig:cylinder_local_errors_probe_2} shows the instantaneous errors ($(\tilde{\varphi} - \varphi)/\varphi^\mathrm{ref} \cdot 100\%$) of reconstructed pressure (left, $\varphi \to p$), horizontal (center, $\varphi \to v_\mathrm{1}$), and vertical (right, $\varphi \to v_\mathrm{2}$) velocity fields at probe 2 in the cylinder's wake (cf. Fig. 2). The results are qualitatively and quantitatively similar to those of integral quantities, i.e., increased $o$- and $\eta^\mathrm{o}$-values provide more accurate model order reductions. In all cases, the error curves rarely exceed a value of one percent, with maximum [minimum] error values in the velocity field for $o=100$ [$o=800$] being in the order of $\mathcal{O}(10^{-1})$ [$\mathcal{O}(10^{-3})$]. The comparatively low error in the pressure field ($\mathcal{O}(10^{-4} - 10^{-6})$) can again be justified numerically, as the pressure field is characterized by minor pressure differences ($\mathcal{O}(10^{1})$) while featuring a reference value ($P = \mathcal{O}(10^{5})$) of large magnitude. Figure \ref{fig:cylinder_prope2_error_mean} provides the time-averaged error results ($\text{mean}((\tilde{\varphi} - \varphi)/\varphi^\mathrm{ref} \cdot 100\%)$) of reduced pressure (left), horizontal (center), and vertical (right) velocity field, respectively. The error curves all show similar slopes; only the most accurate pressure reduction does not significantly improve the transition from $o=400 $ to $o=800$.

\begin{figure}[!ht]
\centering
\iftoggle{tikzExternal}{
\input{./tikz/cylinder/cylinder_prope1_error_mean.tikz}
}{
\includegraphics{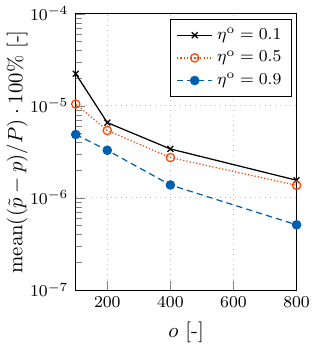}
\includegraphics{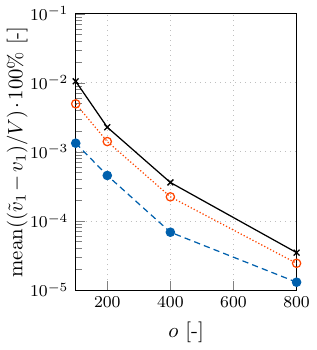}
\includegraphics{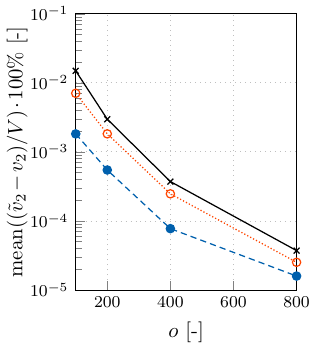}
}
\caption{Cylinder flow ($\mathrm{Re}_\mathrm{D} = 1.4 \cdot 10^5$): Time-averaged reconstruction errors ($\text{mean}((\tilde{\varphi} - \varphi)/\varphi^\mathrm{ref} \cdot 100\%)$) of compressed pressure (left), horizontal (center), and vertical (right) velocity over the minimal construction rank for different retained energy levels $\eta^\mathrm{o}=[0.1,0.5,0.9]$ at probe 1 ($x_\mathrm{1}/D = -1.0$, $x_\mathrm{2}/D = 0.0$, $x_\mathrm{3}/D = 1.0$).}
\label{fig:cylinder_prope1_error_mean}
\end{figure}

\begin{figure}[!ht]
\centering
\iftoggle{tikzExternal}{
\input{./tikz/cylinder/cylinder_prope3_error_mean.tikz}
}{
\includegraphics{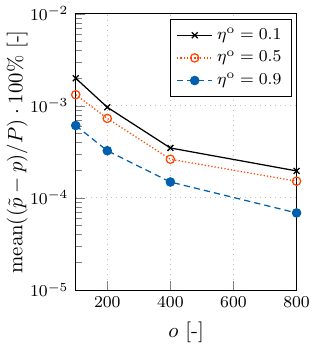}
\includegraphics{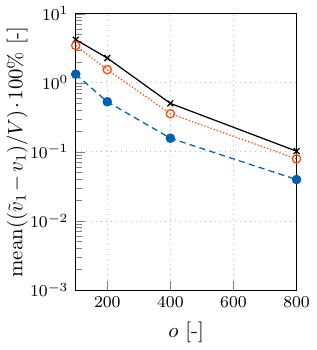}
\includegraphics{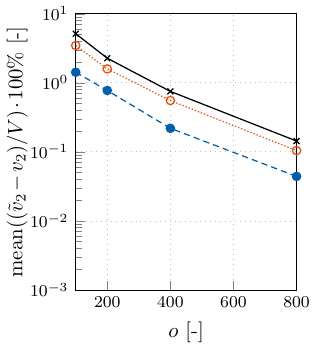}
}
\caption{Cylinder flow ($\mathrm{Re}_\mathrm{D} = 1.4 \cdot 10^5$): Time-averaged reconstruction errors ($\text{mean}((\tilde{\varphi} - \varphi)/\varphi^\mathrm{ref} \cdot 100\%)$) of compressed pressure (left), horizontal (center), and vertical (right) velocity over the minimal construction rank for different retained energy levels $\eta^\mathrm{o}=[0.1,0.5,0.9]$ at probe 3 ($x_\mathrm{1}/D = 1.0$, $x_\mathrm{2}/D = -1.0$, $x_\mathrm{3}/D = 1.0$).}
\label{fig:cylinder_prope3_error_mean}
\end{figure}

\begin{figure}[!ht]
\centering
\iftoggle{tikzExternal}{
\input{./tikz/cylinder/cylinder_prope4_error_mean.tikz}
}{
\includegraphics{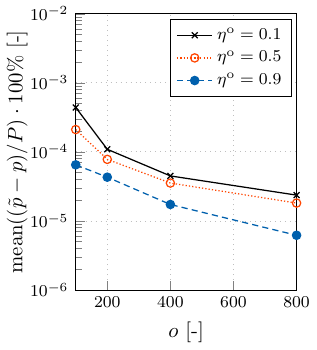}
\includegraphics{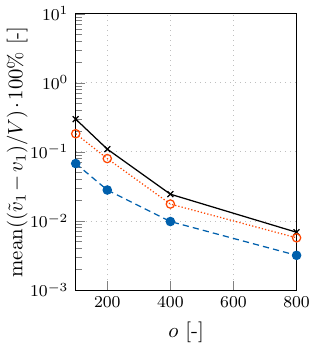}
\includegraphics{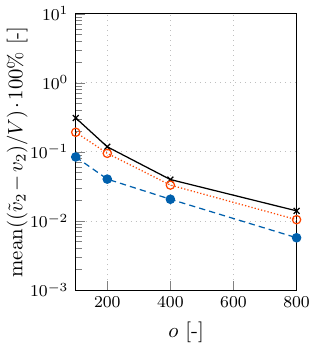}
}
\caption{Cylinder flow ($\mathrm{Re}_\mathrm{D} = 1.4 \cdot 10^5$): Time-averaged reconstruction errors ($\text{mean}((\tilde{\varphi} - \varphi)/\varphi^\mathrm{ref} \cdot 100\%)$) of compressed pressure (left), horizontal (center), and vertical (right) velocity over the minimal construction rank for different retained energy levels $\eta^\mathrm{o}=[0.1,0.5,0.9]$ at probe 4 ($x_\mathrm{1}/D = 5.0$, $x_\mathrm{2}/D = 0.0$, $x_\mathrm{3}/D = 1.0$).}
\label{fig:cylinder_prope4_error_mean}
\end{figure}

Figures \ref{fig:cylinder_prope1_error_mean}-\ref{fig:cylinder_prope4_error_mean} present the time-averaged flow field reconstruction errors of the remaining samples 1, 3, and 4. The results are qualitatively similar to those of probe 2, with quantitative differences in sample 1 [4], where errors tend to be lower [higher]. The decreased reduction errors in probe 1 follow from the spatial positioning upstream of the cylinder with correspondingly few temporal changes. Appendix \ref{app:circular_cylinder} provides the instantaneous errors for $o=[100, 200, 400, 800]$ with $\eta^\mathrm{o} = 0.9$ for completeness, cf. Figs. \ref{fig:cylinder_local_errors_probe_1}-\ref{fig:cylinder_local_errors_probe_4}.

\subsection{Ship Aerodynamics}
\begin{figure}[!ht]
\begin{minipage}[c][11cm][t]{0.4\textwidth}
	\vspace*{\fill}
	\centering
	\subfigure[]{
	\includegraphics[width=0.9\textwidth]{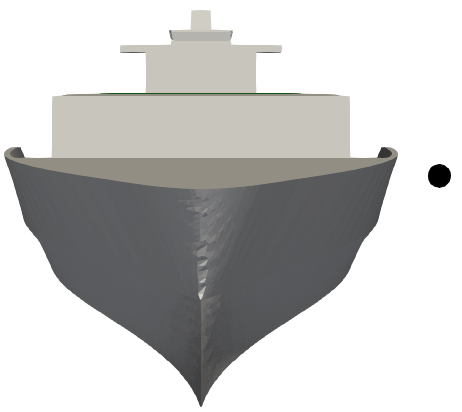}
	}
\end{minipage}
\begin{minipage}[c][11cm][t]{0.6\textwidth}
	\vspace*{\fill}
	\centering
	\subfigure[]{
	\includegraphics[width=1.0\textwidth]{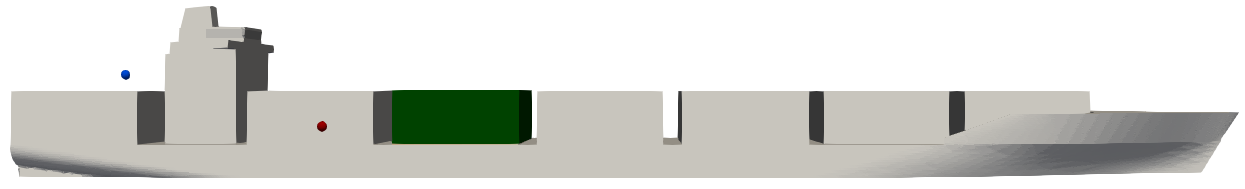}
	}
	\subfigure[]{
	\includegraphics[width=1.0\textwidth]{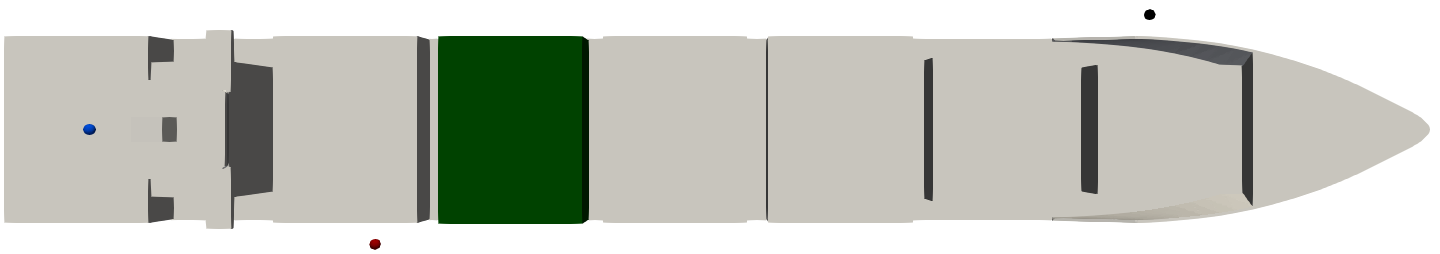}
	}
\end{minipage}
\caption{Feeder ship flow ($\mathrm{Re}_\mathrm{L} = 5.0 \cdot 10^8$): Representation of the geometry from a) front, b) starboard side, and c) top, as well as the indication of the four measuring points used for local evaluation. Additionally, the container to asses integral quantities is highlighted in green.}
\label{fig:kcs_propes}
\end{figure}

\begin{figure}[!ht]
\begin{minipage}[c][11cm][t]{0.3\textwidth}
	\vspace*{\fill}
	\centering
	\subfigure[]{
	\includegraphics[width=0.9\textwidth]{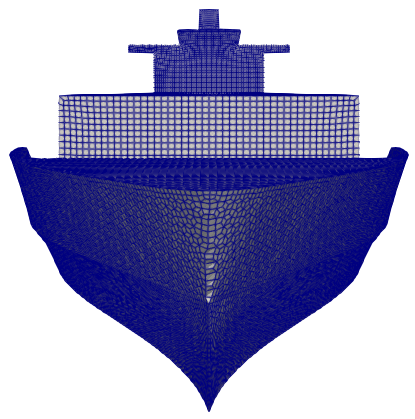}
	}
\end{minipage}
\begin{minipage}[c][11cm][t]{0.7\textwidth}
	\vspace*{\fill}
	\centering
	\subfigure[]{
	\includegraphics[width=1.0\textwidth]{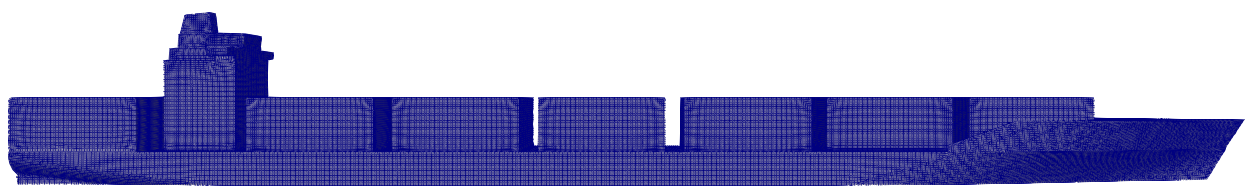}
	}
	\subfigure[]{
	\includegraphics[width=1.0\textwidth]{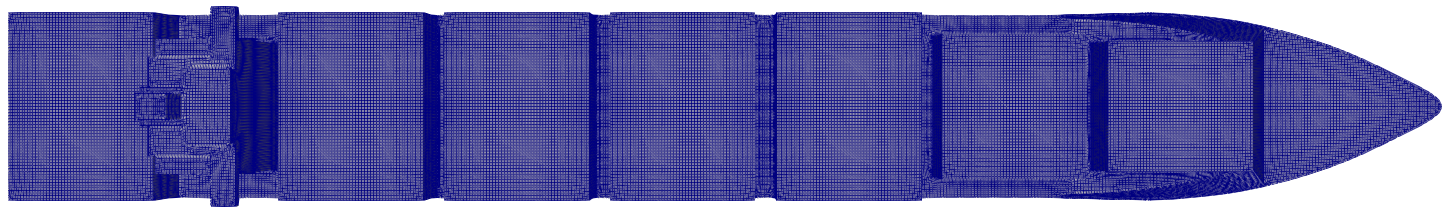}
	}
\end{minipage}
\caption{Feeder ship flow ($\mathrm{Re}_\mathrm{L} = 5.0 \cdot 10^8$): The utilized surface discretization presented from a) front, b) starboard side, and c) top, cf. Fig. \ref{fig:kcs_propes}.}
\label{fig:kcs_grid}
\end{figure}

\begin{figure}[!ht]
\begin{minipage}[c][11cm][t]{1.0\textwidth}
	\vspace*{\fill}
	\centering
	\subfigure[]{
	\includegraphics[width=0.475\textwidth]{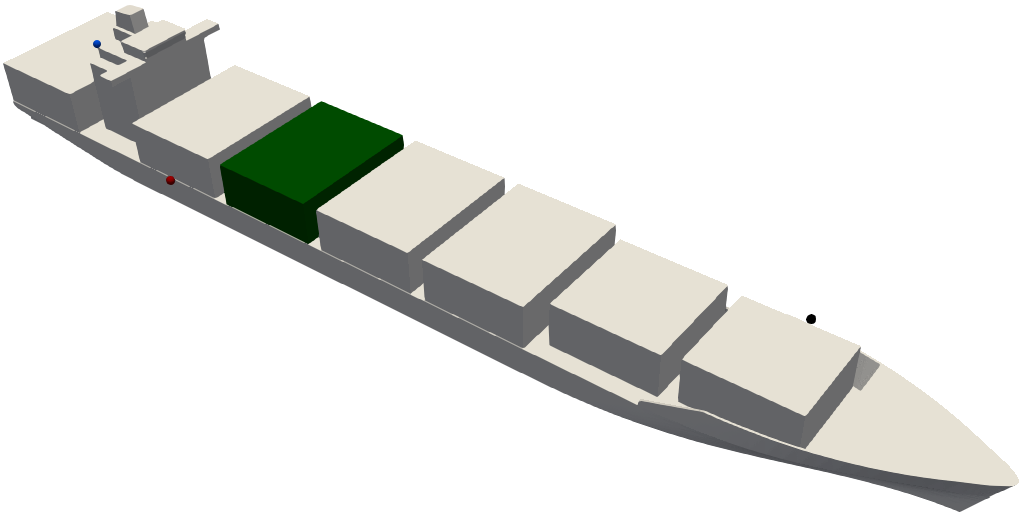}
	}
	\subfigure[]{
	\includegraphics[width=0.475\textwidth]{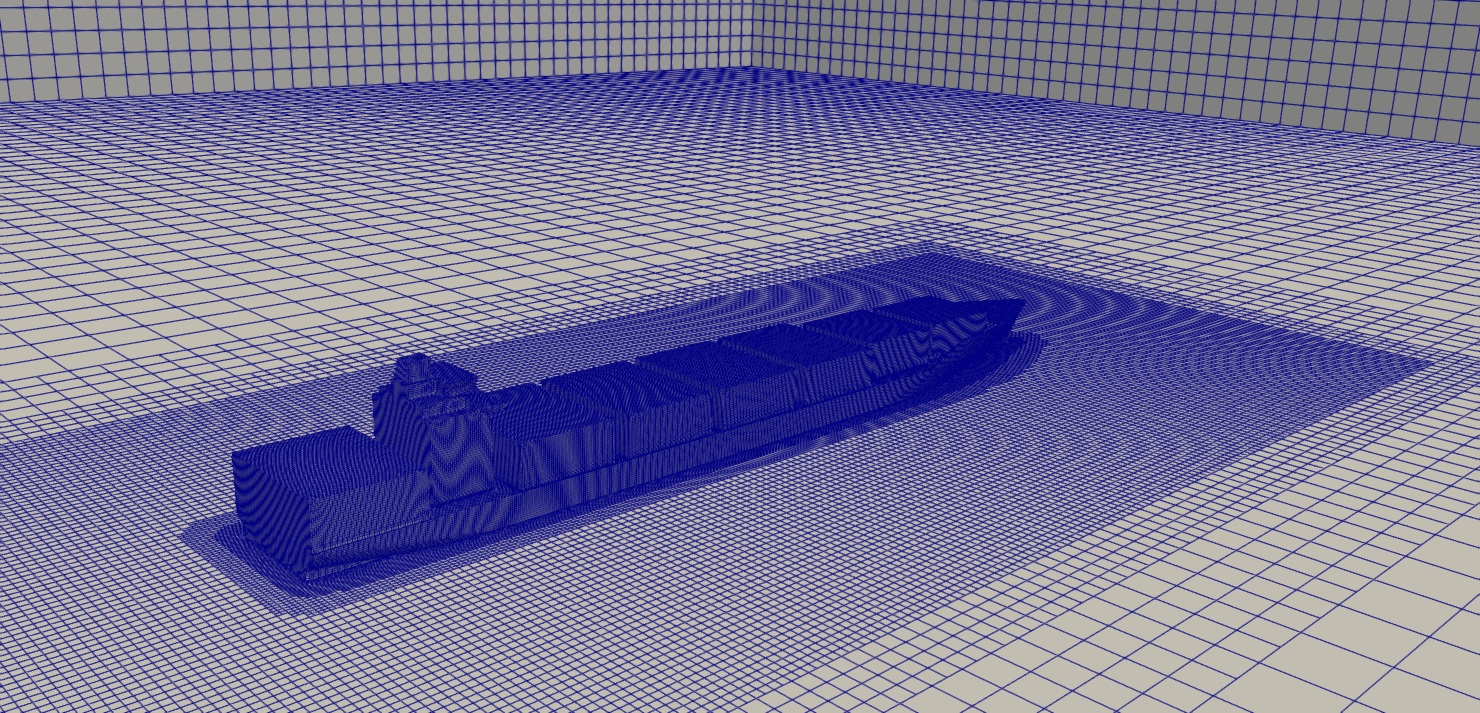}
	}
\end{minipage}
\caption{Feeder ship flow ($\mathrm{Re}_\mathrm{L} = 5.0 \cdot 10^8$): (a) Perspective view of the considered geometry, cf. Fig. \ref{fig:kcs_propes}, and (b) part of the numerical near and far-field grid, cf. Fig. \ref{fig:kcs_grid}.}
\label{fig:kcs_perspectives}
\end{figure}

\begin{figure}[!ht]
\centering
\subfigure[]{
\iftoggle{tikzExternal}{
\input{./tikz/kcs/kcs_vortices.tikz}
}{
\includegraphics{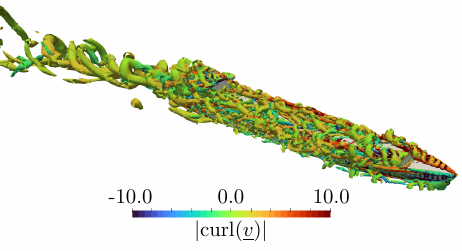}
}
}
\subfigure[]{
\iftoggle{tikzExternal}{
\input{./tikz/kcs/kcs_streamlines.tikz}
}{
\includegraphics{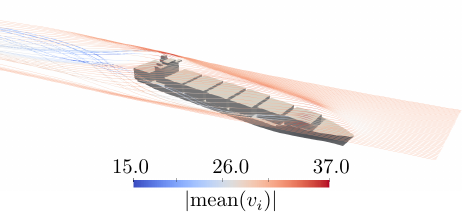}
}
}
\caption{Feeder ship flow ($\mathrm{Re}_\mathrm{L} = 5.0 \cdot 10^8$): Perspective view on (a) isolines of the $\lambda_2$-criterion for $\lambda_2 = -10$ highlighted by the vorticity magnitude and on (b) streamlines of the averaged velocity field seeded on the port-sided front ship colored by the averaged velocities magnitude.}
\label{fig:kcs_results}
\end{figure}

The industrial application example addresses the flow around the full-scale superstructure of a feeder ship equipped with a generic container distribution, cf. Figs. \ref{fig:kcs_propes}-\ref{fig:kcs_perspectives}, as well as \cite{angerbauer2020hybrid, pache2022data, pache2023datenbasierte}, where all further geometry details are provided. The study is conducted at $\mathrm{Re}_\mathrm{L} = V L / \nu = 5.0 \cdot 10^8$ based on the ship's length $L$ and the apparent (air, $\nu$) wind speed $V$ approaching from $20^\circ$ port.
To reduce the blocking effect, the domain extends ten/seven/two ship lengths in the longitudinal/lateral/vertical, i.e., $x_1$/$x_2$/$x_3$ direction, where the superstructure is located centered in $x_1$, $x_2$ direction, and at the minimal $x_3$ coordinate.
A numerical grid with approximately four million unstructured hexahedral control volumes --optimized for the inflow angles of 20 degrees-- is used. In addition to a refinement region in the expected wakefield of the superstructure, the ship is discretized with about 200 isotropic elements along its length. A universal wall function, which blends between the viscous sub-layer and a standard wall function (\cite{rung2001universal, gritskevich2017comprehensive}), is employed with a resolution of $y^+ \approx 50$.
Diffusive fluxes are approximated using a second-order Central Difference Method (CDS) method. The discretization of convective fluxes is composed of a blending between a CDS and a Quadratic Interpolation of Convective Kinematics (QUICK) approach. In order to keep the numerical diffusion low and the numerical stability high, 20\% QUICK and 80\% CDS are used. Together with a second-order accurate ITTL time integration scheme, this results in a numerical method that adequately represents the turbulence decay, as previously shown in \cite{angerbauer2020hybrid, pache2023datenbasierte} based on validation studies against experimental data. A time horizon of $(T \, \Delta t) / (L/V) = 25$ equivalent flows is simulated, whereby the constant discrete time step $\Delta t$ is chosen so that the local cell Courant number has a value well below unity, resulting in a total of $T = 30000$ time steps, cf. Eqn. \eqref{equ:adaptive_energy}. An equivalent ship flow is therefore characterized with approx. 1200 time steps. The tSVD construction does not address the initial transient part and becomes active after approximately eight advective passings, i.e., $t^\mathrm{s}/(L/V) = 8$ and $t^\mathrm{e}/(L/V) = 25$ or $T = \SI{20 000}{}$ in Eqn. \eqref{equ:adaptive_energy}.
Along the outer boundaries, an atmospheric wind profile --$v_i(x_3) = V_i (x_3/x_3^\mathrm{ref})^\alpha$ with $\alpha = 0.11$ and $x_3^\mathrm{ref} = L/8$--, composed of the ship speed and the acting wind, is imposed for the velocity field and turbulent variables. The pressure is specified at the domain's upper end, and a symmetry condition models the (calm) free water surface at the lower end of the domain. Initial conditions follow the far field boundaries’ constant, horizontal wind profile, and hydrostatic pressure distribution.
%
%
%
%
In line with the previous cylinder study, both the minimum tSVD rank $o = [100, 200, 400, 800, 1600]$, as well as the minimum energy to be maintained $\eta^\mathrm{o} \cdot 100\% = [10, 50, 90]$ is varied, resulting in 15 different, adaptively determined ranks $R$. Judgment of the tSVD's qualities follows from integral (drag coefficient) as well as local (velocity, pressure, and Turbulent Kinetic Energy (TKE) as well as its Specific Dissipation Rate (SDR)) flow data, cf. App. \ref{subsec:detached_eddy_simulation} for further flow model details. To account for the local data, three probes are positioned around the Feeder's superstructure, cf. Fig. \ref{fig:kcs_propes}: Two on port/star side in the fore \& aft ship at $x_i/L = [0.8, 0.08, 0.04]^\mathrm{T}$ \& $x_i/L = [0.225, -0.08, 0.04]^\mathrm{T}$, respectively, and a third one on the ship's symmetry plane directly behind the superstructure at $x_i/L = [0.04, 0.0, 0.08]^\mathrm{T}$, measured from the stern center. Global data for the fifth container (counted from the front ship) and the deckhouse are assessed. Each computed field quantity is considered by an isolated itSVD, i.e., four separate tSVD's are constructed for the scalar pressure (turbulent kinetic energy ) [specific dissipation rate] field $p$ ($k$) [$\omega$] and the vector-valued velocity field $v_\mathrm{i}$. Figure \ref{fig:kcs_results} provides flow impression by depicting (a) vortex structures based on the $\lambda_2$ criterion ($\lambda_2 = -10$ colored by the vorticity magnitude) as well as (b) streamlines seeded on the front ship region following the averaged velocity field colored by its magnitude.

First, the different tSVDs are analyzed quantitatively. For this purpose, the numbers of adaptively determined singular values $\mathrm{R}^\mathrm{\varphi}$, cf. Eqn. \eqref{equ:adaptive_energy}, are given in Tab. \ref{tab:kcs_svd_results} for all 15 scenarios of the vector-valued velocity ($\varphi \to v$), and the scalar pressure ($\varphi \to p$), turbulent kinetic energy ($\varphi \to k$), as well as its specific dissipation rate ($\varphi \to \omega$) field. Additionally, the construction time required for each tSVD $t^\mathrm{svd, \varphi}$ (normalized with the total simulation time of the flow solver $t^\mathrm{cfd}$), as well as the required storage $d^\mathrm{svd, \varphi}$ (normalized with the capacity of a full-storage storage approach $d^\mathrm{full, \varphi}$), is provided. Data for the latter follows a general approximation, cf. the rough estimation at the beginning of Sec. \ref{sec:model_order_reduction}.
\begin{table}[!ht]
\caption{Feeder ship flow ($\mathrm{Re}_\mathrm{L} = 5.0 \cdot 10^8$): Number of adaptively computed singular values $\mathrm{R}^\mathrm{\varphi}$ for all equation-determined quantities, i.e., the vector-valued velocity ($\varphi \to v$), and the scalar pressure ($\varphi \to p$), turbulent kinetic energy ($\varphi \to k$), as well as its specific dissipation rate ($\varphi \to \omega$) field. In addition, the relative simulation effort $t^\mathrm{svd, \varphi}$ (normalized with the simulation time of the flow solver $t^\mathrm{cfd}$) and storage overhead $d^\mathrm{svd, \varphi}$ (normalized with the required storage capacity of a full storage approach $d^\mathrm{full, \varphi}$) are provided.}
\begin{center}
\scriptsize
\begin{tabular}{c|ccc|ccc|ccc|ccc|ccc}
\toprule
$o$ & \multicolumn{3}{c|}{100} & \multicolumn{3}{c|}{200} & \multicolumn{3}{c|}{400} & \multicolumn{3}{c|}{800} & \multicolumn{3}{c}{1600} \\ \midrule 
$\eta^\mathrm{o} \cdot 100\%$ & 10 & 50 & 90 & 10 & 50 & 90 & 10 & 50 & 90 & 10 & 50 & 90 & 10 & 50 & 90  \\ \midrule \midrule
$\mathrm{R}^\mathrm{v}$ & 125 & 267 & 680 & 226 & 373 & 799 & 427 & 579 & 1038 & 830 & 1005 & 1537 & 1638 & 1865 & 2260 \\ \midrule 
$t^\mathrm{svd, v}/t^\mathrm{cfd} \cdot 100\%$ & 1.02 & 1.07 & 1.37 & 1.10 & 1.25 & 1.55 & 1.37 & 1.50 & 2.12 & 2.65 & 2.43 & 3.59 & 3.83 & 4.56 & 7.65 \\ \midrule 
$d^\mathrm{svd, v}/d^\mathrm{full, v} \cdot 100\%$ & 0.62 & 1.33 & 3.40 & 1.11 & 1.86 & 4.00 & 2.138 & 2.89 & 5.19 & 4.15 & 5.033 & 7.69 & 8.20 & 9.34 & 11.31  \\ \midrule \midrule
$\mathrm{R}^\mathrm{p}$ & 126 & 285 & 729 & 230 & 398 & 849 & 431 & 599 & 1071 & 831 & 1010 & 1575 & 1644 & 1915 & 2450 \\ \midrule 
$t^\mathrm{svd, p}/t^\mathrm{cfd} \cdot 100\%$ & 0.59 & 0.65 & 0.94 & 0.67 & 0.82 & 1.10 & 0.91 & 1.04 & 1.64 & 2.05 & 1.87 & 3.15 & 3.10 & 3.89 & 7.64 \\ \midrule 
$d^\mathrm{svd, p}/d^\mathrm{full, p} \cdot 100\%$ & 0.63 &1.43 & 3.66 & 1.15 & 1.95 & 4.26 & 2.16 & 3.01 & 5.38 & 4.17 & 5.07 & 7.91 & 8.26 & 9.62 & 12.31\\ \midrule \midrule
$\mathrm{R}^\mathrm{k}$ & 131 & 291 & 836 & 234 & 420 & 955 & 434 & 624 & 1189 & 836 & 1046 & 1683 & 1645 & 1904 & 2342 \\ \midrule 
$t^\mathrm{svd, k}/t^\mathrm{cfd} \cdot 100\%$ & 0.59 & 0.58 & 1.01 & 0.67 & 0.84 & 1.18 & 0.92 & 1.05 & 1.72 & 2.05 & 1.88 & 3.03 & 3.07 & 3.74 & 6.43 \\ \midrule 
$d^\mathrm{svd, k}/d^\mathrm{full, k} \cdot 100\%$ & 0.65 & 1.46 & 4.20 & 1.17 & 2.11 & 4.79 & 2.18 & 3.13 & 5.97 & 4.20 & 5.25 & 8.45 & 8.26 & 9.56 & 11.76 \\ \midrule \midrule
$\mathrm{R}^\mathrm{\omega}$ & 127 & 282 & 738 & 229 & 391 & 863 & 430 & 599 & 1109 & 833 & 1028 & 1628 & 1644 & 1904 & 2361 \\ \midrule 
$t^\mathrm{svd, \omega}/t^\mathrm{cfd} \cdot 100\%$ & 0.59 & 0.64 & 0.95 & 0.67 & 0.82 & 1.12 & 0.92 & 1.03 & 1.67 & 2.05 & 1.86 & 3.11 & 3.11 & 3.84 & 7.05 \\ \midrule 
$d^\mathrm{svd, \omega}/d^\mathrm{full, \omega} \cdot 100\%$ & 0.63 & 1.41 & 3.70 & 1.15 & 1.96 & 4.33 & 2.16 & 3.01 & 5.57 & 4.18 & 5.16 & 8.18 & 8.26 & 9.56 & 11.86 \\ \midrule 
\bottomrule
\end{tabular}
\end{center}
\label{tab:kcs_svd_results}
\end{table}

\begin{figure}[!ht]
\centering
\iftoggle{tikzExternal}{
\input{./tikz/kcs/kcs_singular_value_adaption.tikz}
}{
\includegraphics{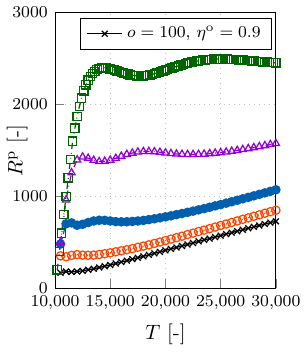}
\includegraphics{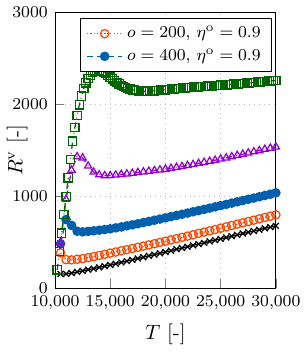}
\includegraphics{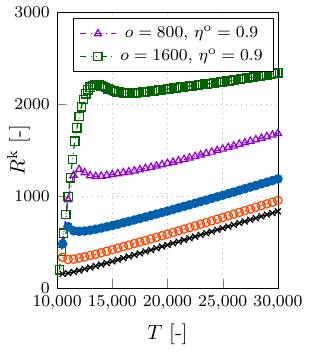}
}
\caption{Feeder ship flow ($\mathrm{Re}_\mathrm{L} = 5.0 \cdot 10^8$): Number of considered singular values of the pressure (left, $R^\mathrm{p}$), velocity (center, $R^\mathrm{v}$), and turbulent kinetic energy (right, $R^\mathrm{k}$) field over the simulated time steps for minimal construction ranks $o=[100, 200, 400, 800]$ with a retained energy of $\eta^\mathrm{o} = 0.9$ each.}
\label{fig:kcs_singular_value_adaption}
\end{figure}

The tSVD data of the four different field quantities are quantitatively and qualitatively similar. As expected, the number of singular values reach their minimum [maximum] for the lowest [highest] minimum construction ranks ($o=100$) [($o=1600$)], as well as the smallest ($\eta^\mathrm{o} = 0.1$) [largest ($\eta^\mathrm{o} = 0.9$)] energy to be retained. Figure \ref{fig:kcs_singular_value_adaption} underlines this by providing the adaptively determined number of singular values of the pressure (left, $R^\mathrm{p}$), the velocity field (center, $R^\mathrm{v}$), and the turbulent kinetic energy (right, $R^\mathrm{k}$) over the number of the itSVD-relevant time steps for all considered minimum construction ranks $o=[100, 200, 400, 800, 1600]$ and a retained energy level of $\eta^\mathrm{o} = 0.9$. Compared to the cylinder study (see Fig. \ref{fig:cylinder_singular_value_adaption}), a less abrupt increase in singular values arises over the simulation time. After an initial overshoot between $10000 \leq T \leq 15000$, the amount of singular values constantly increases from $T>20000$, and the gradient is the highest [lowest] for the lowest [highest] minimum construction rank. The number of singular pressure values shows a somewhat more volatile course.
Further observations align with results reported for the LES case from the previous Sec. \ref{subsec:cylinder}. The remaining thirteen scenarios yield values between the two extreme cases, where compared to higher energy constraints, increased lower rank limits tend to result in an increased number of singular values. Again, the simulation and storage overheads scale with the number of singular values considered, with a time overhead of around $\mathcal{O}(1)$\% for a small number of singular values and about $\approx 7$\% for the higher reduction cases. Besides, the storage overhead varies between $\mathcal{O}(1)$\% and $\mathcal{O}(10)$\%.  Again, note that the provided numbers are only indicative since compression effects due to the employed H5 library and hardware (inter-processor communication, etc.), as well as software (algorithms, etc.) related aspects, could impair the results. 

\begin{figure}[!ht]
\centering
\iftoggle{tikzExternal}{
\input{./tikz/kcs/kcs_all_f1.tikz}
}{
\includegraphics{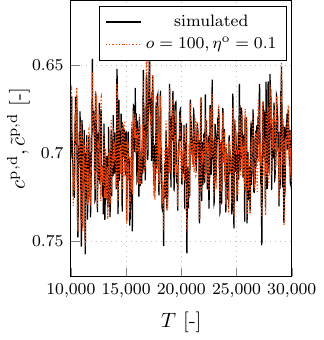}
\includegraphics{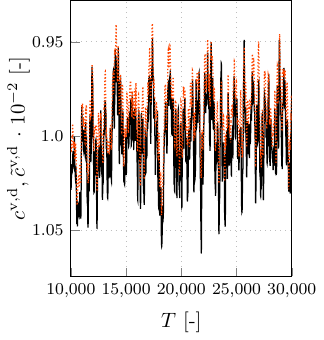}
\includegraphics{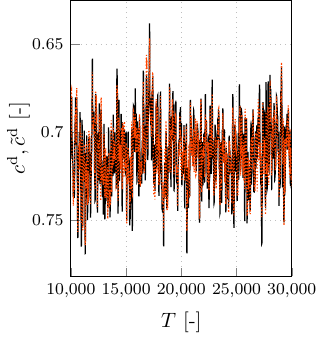}
}
\caption{Feeder ship flow ($\mathrm{Re}_\mathrm{L} = 5.0 \cdot 10^8$); Entire superstructure: Simulated and reconstructed pressure (left), viscous (center), and total (right) drag coefficient over the simulated time steps.}
\label{fig:kcs_all_forces}
\end{figure}

In line with the assessment of the LES results from the previous Sec. \ref{subsec:cylinder}, results of several reduced order models are considered in the following based on a re-evaluation of the compressed flow fields while comparing against the simulated reference data. Global data is assessed based on the drag coefficient only since the results of lateral force coefficients are similar and, therefore, not included to save space. Figure \ref{fig:kcs_all_forces} depicts the simulated and (with most minor parameters, i.e., $o=100, \eta^\mathrm{o}=0.1$) reconstructed pressure (left), viscous (center), and total (right) drag coefficient over the simulated time steps. Even for this small reconstruction rank (see Tab. \ref{tab:kcs_svd_results}), the reconstructed data does not deviate significantly from the simulated solution of the pressure and total resistance coefficient. Only the reconstructed friction component is slightly shifted horizontally. For higher reconstruction ranks, differences are hardly recognizable, so the deviation between the simulated and reconstructed solution is considered directly in the following.

\begin{figure}[!ht]
\centering
\iftoggle{tikzExternal}{
\input{./tikz/kcs/kcs_all_error_f1.tikz}
}{
\includegraphics{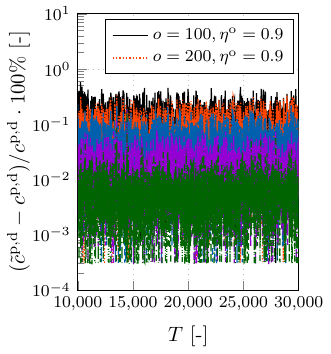}
\includegraphics{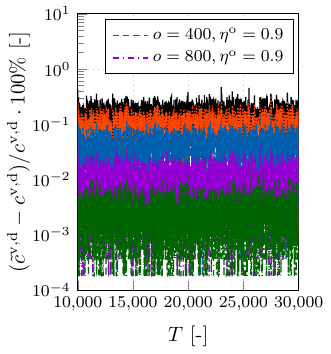}
\includegraphics{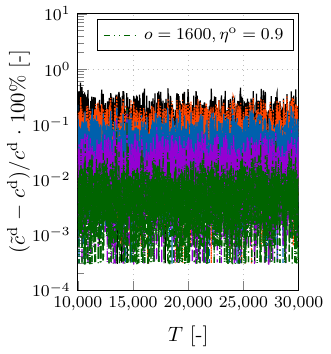}
}
\caption{Feeder ship flow ($\mathrm{Re}_\mathrm{L} = 5.0 \cdot 10^8$); Entire superstructure: Instantaneous relative error of pressure (left), viscous (center), and total (right) drag coefficient over the simulated time steps T for varying minimal tSVD ranks $o = [100, 200, 400, 800, 1600]$ for a retained energy level of $\eta^\mathrm{o} = 0.9$.}
\label{fig:kcs_all_error_f1}
\end{figure}

The instantaneous error $(\tilde{c}^{(\cdot, d)} - c^{(\cdot, d)})/c^{(\cdot, d)} \cdot 100\%$) of the reconstructed pressure ($\tilde{c}^\mathrm{p,d}$, left), viscous ($\tilde{c}^\mathrm{v,d}$, center), and total ($\tilde{c}^\mathrm{d}$, right) drag coefficient of the entire superstructure is shown Fig. \ref{fig:kcs_all_error_f1} for the five considered minimum construction ranks $o=[100, 200, 400, 800, 1600]$ with a maximum retained energy level of $\eta^\mathrm{o} = 0.9$. Due to the increased number of considered singular values for higher minimum construction ranks (cf. Tab. \ref{tab:kcs_svd_results}), the errors decrease from $\mathcal{O}(10^{-1})\%$ to $\mathcal{O}(10^{-2})\%$ for higher $o$ values. The latter is underlined in Fig. \ref{fig:cylinder_integral_error_mean}, where the time-averaged reconstruction errors are plotted over the minimum construction rank for all fifteen considered cases, i.e., with varied retained energy levels. 
\begin{figure}[!ht]
\centering
\iftoggle{tikzExternal}{
\input{./tikz/kcs/kcs_all_mean_error_f1.tikz}
}{
\includegraphics{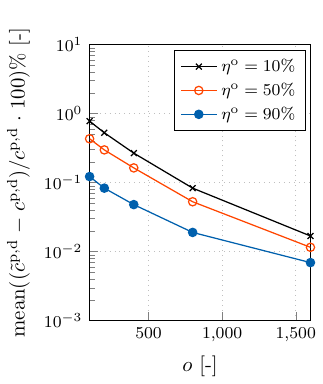}
\includegraphics{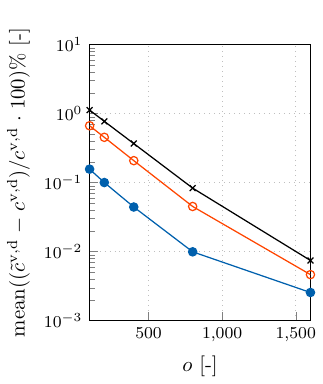}
\includegraphics{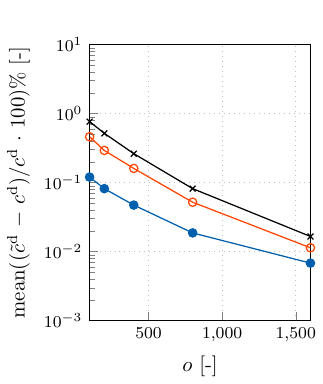}
}
\caption{Feeder ship flow ($\mathrm{Re}_\mathrm{L} = 5.0 \cdot 10^8$); Entire superstructure: Averaged relative error of pressure (left), viscous (center), and total (right) drag coefficient over the varying minimal tSVD ranks $o = [100, 200, 400, 800, 1600]$ for different retained energy levels $\eta^\mathrm{o} \cdot 100\% = [10, 50, 90]$.}
\label{fig:kcs_all_mean_error_f1}
\end{figure}

\begin{figure}[!ht]
\centering
\iftoggle{tikzExternal}{
\input{./tikz/kcs/kcs_container_mean_error_f1.tikz}
}{
\includegraphics{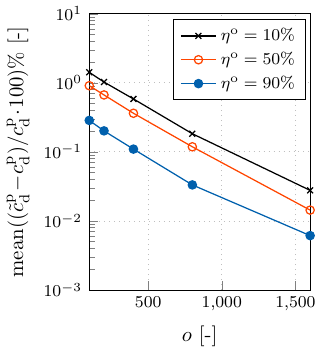}
\includegraphics{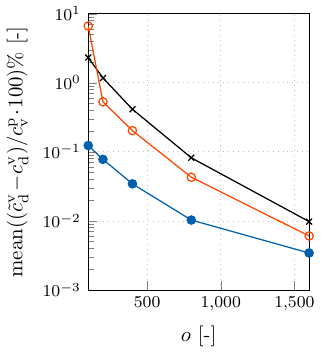}
\includegraphics{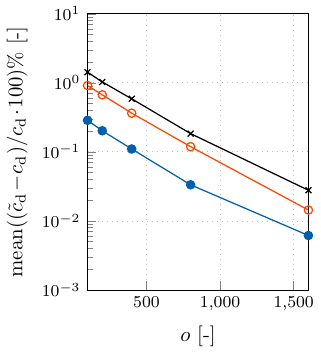}
}
\caption{Feeder ship flow ($\mathrm{Re}_\mathrm{L} = 5.0 \cdot 10^8$); Fifth container: Averaged relative error of pressure (left), viscous (center), and total (right) drag coefficient over the varying minimal tSVD ranks $o = [100, 200, 400, 800, 1600]$ for different retained energy levels $\eta^\mathrm{o} \cdot 100\% = [10, 50, 90]$.}
\label{fig:kcs_container_mean_error_f1}
\end{figure}

\begin{figure}[!ht]
\centering
\iftoggle{tikzExternal}{
\input{./tikz/kcs/kcs_deckshouse_mean_error_f1.tikz}
}{
\includegraphics{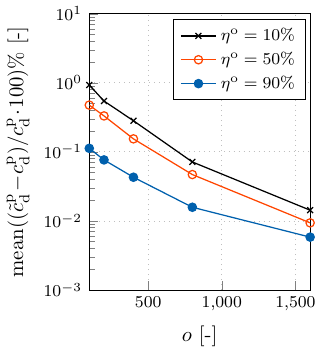}
\includegraphics{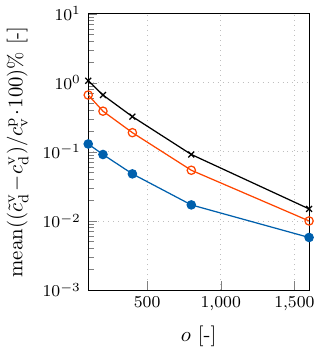}
\includegraphics{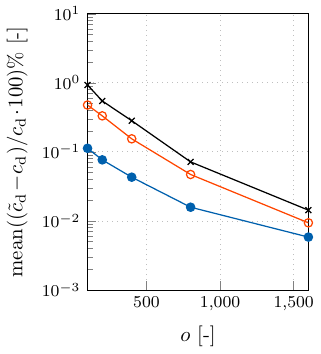}
}
\caption{Feeder ship flow ($\mathrm{Re}_\mathrm{L} = 5.0 \cdot 10^8$); Deckhouse: Averaged relative error of pressure (left), viscous (center), and total (right) drag coefficient over the varying minimal tSVD ranks $o = [100, 200, 400, 800, 1600]$ for different retained energy levels $\eta^\mathrm{o} \cdot 100\% = [10, 50, 90]$.}
\label{fig:kcs_deckshouse_mean_error_f1}
\end{figure}

The evaluation is repeated for the drag coefficient of the fifth container as well as the deckhouse (cf. Figs. \ref{fig:kcs_propes}, \ref{fig:kcs_perspectives}), and corresponding mean results are provided in Figs. \ref{fig:kcs_container_mean_error_f1}-\ref{fig:kcs_deckshouse_mean_error_f1}. The results are quantitatively and qualitatively similar to the previously presented mean results, although the errors for for forces acting on the container are slightly increased for $\eta^\mathrm{o} = [0.1, 0.5]$. The observations of the previous LES studies are repeated in that the mean relative error is $\mathcal{O}(10^{-1})$ when the minimum construction rank is chosen in line with the number of discrete time steps necessary to conduct for one equivalent flow, i.e., $o \approx 1200$ here.
The instantaneous data for $\eta^\mathrm{o} = 0.9$ is shifted to App. \ref{app:fleeder_ship} to keep the presentation concise, i.e., Figs. \ref{fig:kcs_container_error_f1}-\ref{fig:kcs_deckhouse_error_f1}. 

\begin{figure}[!ht]
\centering
\iftoggle{tikzExternal}{
\input{./tikz/kcs/kcs_prope_center_1.tikz}
}{
\includegraphics{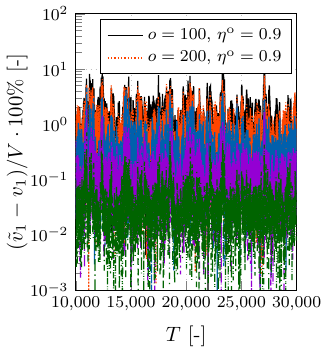}
\includegraphics{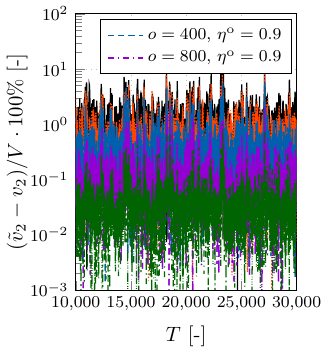}
\includegraphics{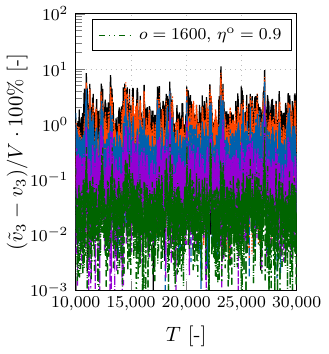}
}
\caption{Feeder ship flow ($\mathrm{Re}_\mathrm{L} = 5.0 \cdot 10^8$); Centered probe: Instantaneous error of longitudinal (left), lateral (center), and vertical (right) velocity over the varying minimal tSVD ranks $o = [100, 200, 400, 800, 1600]$ for different retained energy levels $\eta^\mathrm{o} \cdot 100\% = [10, 50, 90]$.}
\label{fig:kcs_prope_center_error_1}
\end{figure}

\begin{figure}[!ht]
\centering
\iftoggle{tikzExternal}{
\input{./tikz/kcs/kcs_prope_center_2.tikz}
}{
\includegraphics{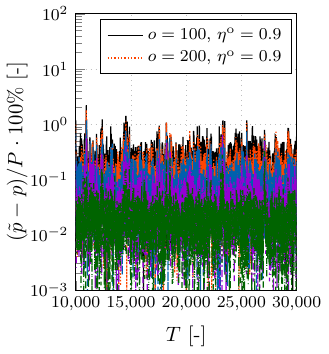}
\includegraphics{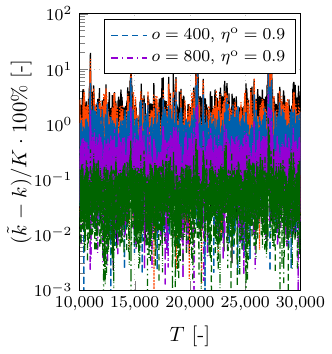}
\includegraphics{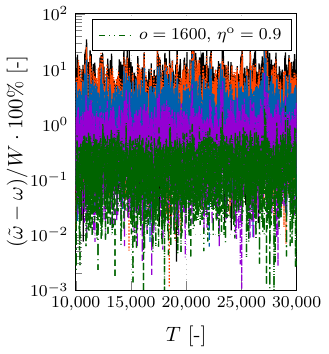}
}
\caption{Feeder ship flow ($\mathrm{Re}_\mathrm{L} = 5.0 \cdot 10^8$); Centered probe: Instantaneous error of pressure (left), TKE (center), and SDR (right) over the varying minimal tSVD ranks $o = [100, 200, 400, 800, 1600]$ for different retained energy levels $\eta^\mathrm{o} \cdot 100\% = [10, 50, 90]$.}
\label{fig:kcs_prope_center_error_2}
\end{figure}

\begin{figure}[!ht]
\centering
\iftoggle{tikzExternal}{
\input{./tikz/kcs/kcs_prope_center_1_mean_error.tikz}
}{
\includegraphics{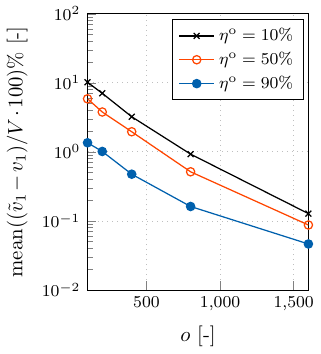}
\includegraphics{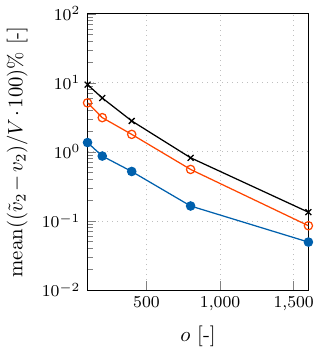}
\includegraphics{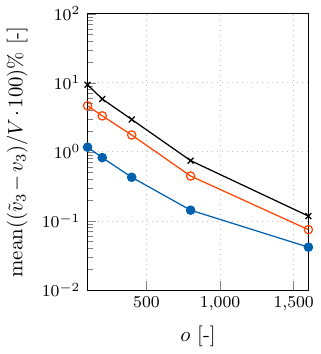}
}
\caption{Feeder ship flow ($\mathrm{Re}_\mathrm{L} = 5.0 \cdot 10^8$); Centered probe: Averaged error of longitudinal (left), lateral (center), and vertical (right) velocity over the varying minimal tSVD ranks $o = [100, 200, 400, 800, 1600]$ for different retained energy levels $\eta^\mathrm{o} \cdot 100\% = [10, 50, 90]$.}
\label{fig:kcs_prope_center_1_mean_error}
\end{figure}

\begin{figure}[!ht]
\centering
\iftoggle{tikzExternal}{
\input{./tikz/kcs/kcs_prope_center_2_mean_error.tikz}
}{
\includegraphics{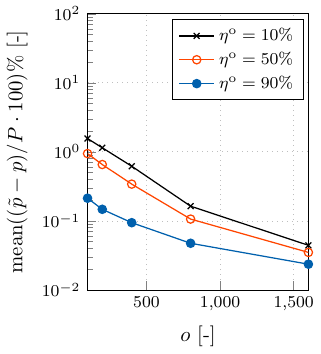}
\includegraphics{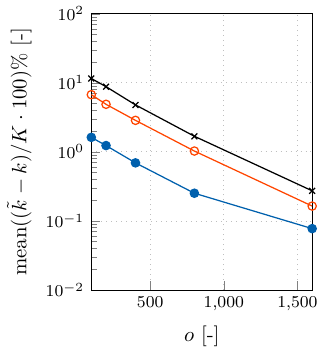}
\includegraphics{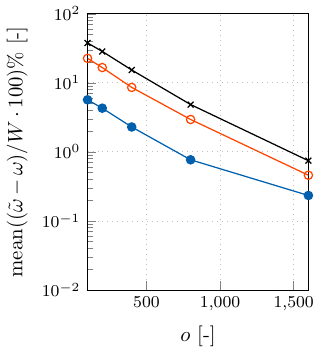}
}
\caption{Feeder ship flow ($\mathrm{Re}_\mathrm{L} = 5.0 \cdot 10^8$); Centered probe: Averaged error of pressure (left), TKE (center), and SDR (right) over the varying minimal tSVD ranks $o = [100, 200, 400, 800, 1600]$ for different retained energy levels $\eta^\mathrm{o} \cdot 100\% = [10, 50, 90]$.}
\label{fig:kcs_prope_center_2_mean_error}
\end{figure}

In addition to the previous consideration of integral variables, the focus in the following is now on the condensation of local, instantaneous flow variables. At the three evaluation positions (see this case's introductory Sec. and Fig. \ref{fig:kcs_propes}), all field quantities (velocity field $v_i$, pressure $p$, turbulent kinetic energy $k$, and its specific dissipation rate $\omega$) --obtained from approximating the partial differential Eqns. \eqref{equ:mass_conservation}-\eqref{equ:momentum_conservation}, \eqref{equ:tke_conservation}-\eqref{equ:omega_conservation}-- are considered. Since relativization with field quantities is susceptible to singularities in the latter, spatially/temporally constant reference values were used as an alternative for normalization. Next to the already introduced reference velocity $V$, the error evaluation employs $P = \rho V^2 / 2$, $K = {V^\mathrm{\tau}}^2$, and $W = {V^\mathrm{\tau}}^4$ as reference values for the pressure, turbulent kinetic energy, and its specific dissipation rate, respectively. The chosen friction velocity $V^\mathrm{\tau}$ follows empirical ﬂat plate relations, i.e., $V^\mathrm{\tau} = \tau^\mathrm{w} / \rho$ with $\tau^\mathrm{w} = c^\mathrm{f} \rho V^2 / 2$ and $c^\mathrm{f} = 0.026 / \mathrm{Re}_\mathrm{L}^{1/7}$. The instantaneous errors of all six field quantities evaluated at the centered probe in the stern region are shown in Figs. \ref{fig:kcs_prope_center_error_1}-\ref{fig:kcs_prope_center_error_2} for different minimum construction ranks at a retained energy level of $\eta^\mathrm{o} \cdot 100 = 90$ percent. As expected based on the tabulated values, the error decreases on average for increasing minimum adaption ranks due to the corresponding increasing number of singular values. The decreasing averaged error characteristics can also be concluded from Figs. \ref{fig:kcs_prope_center_1_mean_error}-\ref{fig:kcs_prope_center_2_mean_error}, which show the time-averaged errors of the instantaneous evaluation for all 15 scenarios considered above the prescribed minimum number of $o$ singular values. In line with the paper's LES results on the circular cylinder from Sec. \ref{subsec:cylinder} and the Feeder's integral investigations (cf. Fig. \ref{fig:kcs_all_error_f1}) in this chapter, a reduction in error can be seen for high values of $o$ or $\eta^\mathrm{o}$. Whereas the decrease is qualitatively similar, a quantitative difference in an increased error amplitude of approximately one order of magnitude arises that, however, might be due to the already discussed alternative normalization. Still, the errors are in the order of $\mathcal{O}(10^{-1})$ for a minimal number of singular values aligned with the number of time steps required for one advective passing.

The following discusses only the time-averaged errors on the two remaining samples on the port and starboard sides. Instantaneous results in this regard are provided in App. \ref{app:fleeder_ship}, i.e., Figs. \ref{fig:kcs_prope_star_error_1}-\ref{fig:kcs_prope_port_error_2}. For the starboard side sample, the averaged errors of, e.g., the longitudinal velocity field $v_1$ are both quantitative and qualitative, similar to the previous results of the probe in the aft ship, i.e., for $o=100$, they are of the order of $\mathcal{O}(10^1)$ [$\mathcal{O}(10^0)$] for $\eta^\mathrm{o} = 0.1$ [$\eta^\mathrm{o} = 0.9$], and drop to $\mathcal{O}(10^0)$ [$\mathcal{O}(10^{-1})$] for $o=800$, whereby the errors of the starboard side sample tend to be slightly higher. The situation on the port side is significantly different, as shown in Figs. \ref{fig:kcs_prope_port_1_mean_error}-\ref{fig:kcs_prope_port_2_mean_error}. Here, the averaged errors are generally around two orders of magnitude lower due to the positioning of the sample upstream of the ship's superstructure, which follows from a flow field hitting the ship that contains much less intense turbulence. The associated reduction of local turbulent flow fluctuations leads to more similar snapshot vectors (cf. Fig. \ref{fig:truncated_parallel_rpts_svd}), which are recognized early by the tSVD procedure. It, therefore, leads to increased reduction accuracies due to the nevertheless required minimum design ranks in line with the probe data at more delicate locations, e.g., in the superstructure's wake. Only the error in the pressure field features a less significant slope.
\begin{figure}[!ht]
\centering
\iftoggle{tikzExternal}{
\input{./tikz/kcs/kcs_prope_star_1_mean_error.tikz}
}{
\includegraphics{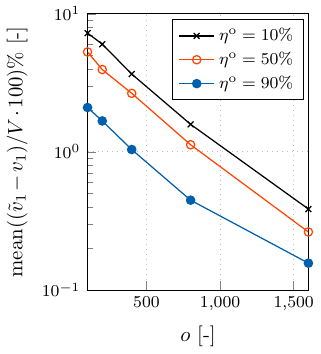}
\includegraphics{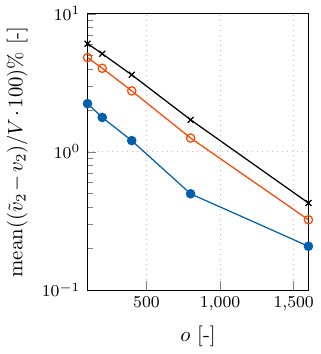}
\includegraphics{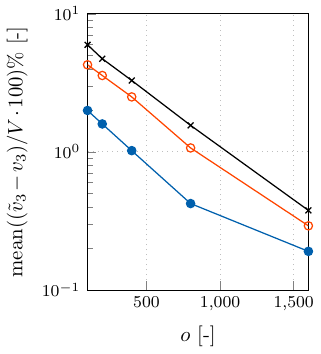}
}
\caption{Feeder ship flow ($\mathrm{Re}_\mathrm{L} = 5.0 \cdot 10^8$); Star side probe: Averaged error of longitudinal (left), lateral (center), and vertical (right) velocity over the varying minimal tSVD ranks $o = [100, 200, 400, 800, 1600]$ for different retained energy levels $\eta^\mathrm{o} \cdot 100\% = [10, 50, 90]$.}
\label{fig:kcs_prope_star_1_mean_error}
\end{figure}
\begin{figure}[!ht]
\centering
\iftoggle{tikzExternal}{
\input{./tikz/kcs/kcs_prope_star_2_mean_error.tikz}
}{
\includegraphics{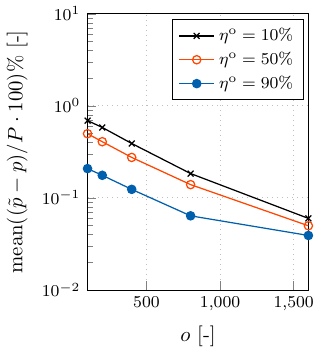}
\includegraphics{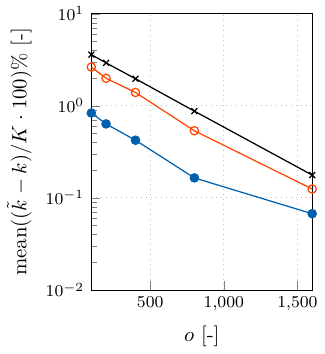}
\includegraphics{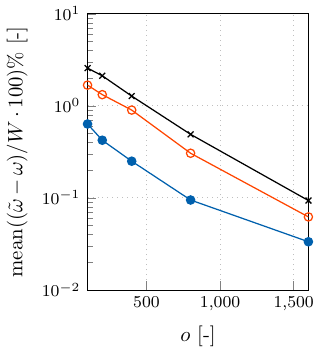}
}
\caption{Feeder ship flow ($\mathrm{Re}_\mathrm{L} = 5.0 \cdot 10^8$); Star side probe: Averaged error of pressure (left), TKE (center), and SDR (right) over the varying minimal tSVD ranks $o = [100, 200, 400, 800, 1600]$ for different retained energy levels $\eta^\mathrm{o} \cdot 100\% = [10, 50, 90]$.}
\label{fig:kcs_prope_star_2_mean_error}
\end{figure}
\begin{figure}[!ht]
\centering
\iftoggle{tikzExternal}{
\input{./tikz/kcs/kcs_prope_port_1_mean_error.tikz}
}{
\includegraphics{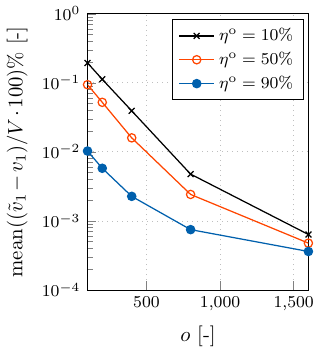}
\includegraphics{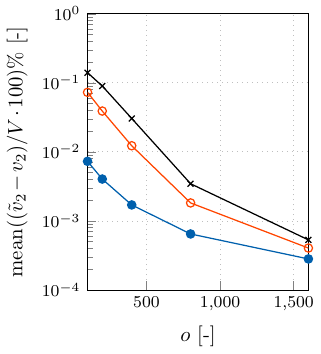}
\includegraphics{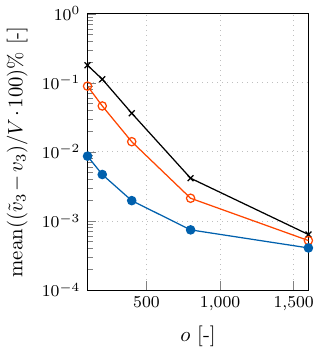}
}
\caption{Feeder ship flow ($\mathrm{Re}_\mathrm{L} = 5.0 \cdot 10^8$); Port side probe: Averaged error of longitudinal (left), lateral (center), and vertical (right) velocity over the varying minimal tSVD ranks $o = [100, 200, 400, 800, 1600]$ for different retained energy levels $\eta^\mathrm{o} \cdot 100\% = [10, 50, 90]$.}
\label{fig:kcs_prope_port_1_mean_error}
\end{figure}
\begin{figure}[!ht]
\centering
\iftoggle{tikzExternal}{
\input{./tikz/kcs/kcs_prope_port_2_mean_error.tikz}
}{
\includegraphics{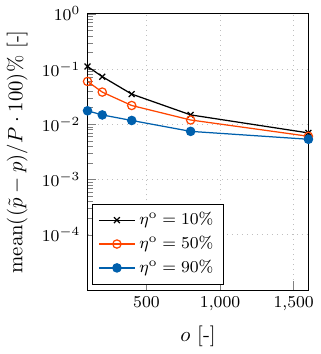}
\includegraphics{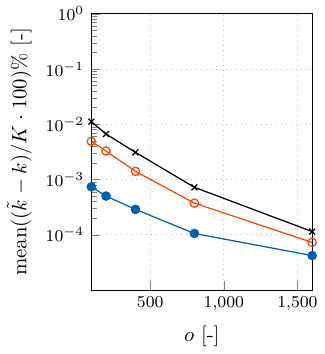}
\includegraphics{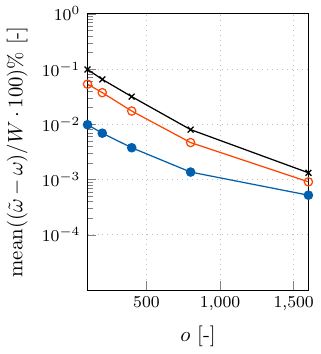}
}
\caption{Feeder ship flow ($\mathrm{Re}_\mathrm{L} = 5.0 \cdot 10^8$); Port side probe: Averaged error of pressure (left), TKE (center), and SDR (right) over the varying minimal tSVD ranks $o = [100, 200, 400, 800, 1600]$ for different retained energy levels $\eta^\mathrm{o} \cdot 100\% = [10, 50, 90]$.}
\label{fig:kcs_prope_port_2_mean_error}
\end{figure}

\section{Conclusion \& Outlook}
The paper applied an incremental truncated Singular Value Decomposition (itSVD) approach to construct a Reduced Order Model (ROM) of scale-resolving Computational Fluid Dynamics (CFD) applications in a fully adaptive, simulation-accompanying approach. Two turbulence modeling strategies have been pursued that refer to (a) an academic Large Eddy Simulation study around a circular cylinder and (b) an industrial scenario that considers the aerodynamics of a generic container ship based on a Detached Eddy Simulation approach.

Results indicate the possibility of assessing redundancy in the resolved instantaneous turbulent fluctuations that the ROM detects which then leads to high compression rates of $\mathcal{O}(95)$\% accompanied by relative errors in the order of $\mathcal{O}(10^{-3}-10^{-1})$ percent, i.e., well below common modeling and discretization errors. The computational overhead was around $\mathcal{O}(10)$\% of the flow solver's simulation time.

To obtain a fully adaptive ROM construction process, two corresponding parameters of practically importance --to be prescribed at simulation start-- have been studied, that refer to (a) the ROM's minimum construction rank as well as (b) the retained energy of the remaining modes. To obtain acceptable results, it was found that the value for
\begin{itemize}
    \item [(a)] 
        should be aligned with the number of discrete time steps needed for one equivalent flow, e.g., around the cylinder or the ship's superstructure. A value, that each CFD user usually knows --at least roughly-- a-priori.
    \item [(b)]
        allows for flexibility, but lower/upper limits of 50/90\% have been shown acceptable results. Choosing values above 99\% would force the adaptive tSVD procedure to compute almost all singular values, which impedes the ROM idea. Hence, a value of 75\% should serve as best-practice.
\end{itemize}

In order to minimize the computational overhead, future work should improve the algorithmic efficiency of the employed itSVD approach, e.g., by accelerated QR decomposition strategies and adaptive decision-making on orthogonalization as well as subspace rotation steps. Further technical aspects could address the application to scale resolving simulations on adaptively refined/coarsened numerical grids. Besides, the application to fully resolved, i.e., Direct Numerical Simulation approaches, would possibly provide further insights into appropriate ROM rank determinations.

\section{Declaration of Competing Interest}
The author declares that he has no known competing financial interests or personal relationships that could have appeared to influence the work reported in this paper.

\section{Acknowledgments}
The current work is part of the “Propulsion Optimization of Ships and Appendages” research project funded by the German Federal Ministry for Economics and Climate Action (Grant No. 03SX599C). The author gratefully acknowledges this support.
In addition, the author would like to thank Dr.-Ing. Rupert Pache for providing geometry and operational data of the surface vessel as well as related best practice flow solver settings.
%
%

\begin{appendix}

\section{Governing Equations}
\label{app:governing_equations}

The conservation of mass and linear momentum governs the temporal ($t$) evolution of an incompressible fluid's (density $\rho$, dynamic viscosity $\mu$) velocity $v_i$ and pressure $p$, viz.
\begin{align}
	\frac{\partial v_\mathrm{k}}{\partial x_\mathrm{k}} &= 0 \label{equ:mass_conservation} \\
	\frac{\partial \rho v_i}{\partial t} + \frac{\partial}{\partial x_k} \left[ v_k \rho v_i + p^\mathrm{eff} \delta_{ik} - 2 \mu^\mathrm{eff} S_{ik} \right] &= 0 \label{equ:momentum_conservation}
\end{align}
where $\delta_{ik}$ denotes the Kronecker delta and $\mu^\mathrm{eff} = \mu + \mu^\mathrm{t}$ as well as $p^\mathrm{eff} = p + p^\mathrm{t}$ refer to effective viscosity and pressure, respectively, that contain turbulent contributions that depend on the underlying flow modeling approach. For this paper's studies, the employed numerical strategies differ in the determination of the eddy viscosity only, which is briefly described in the following.

\subsection{Large Eddy Simulation}
\label{subsec:large_eddy_simulation}

In line with the  Wall-Adapting Local Eddy Viscosity (WALE) model of \cite{ducros1998wall, nicoud1999subgrid}, the turbulent viscosity is expressed as
\begin{align}
    \mu^\mathrm{t} &= \rho {l^\mathrm{s}}^2 \frac{[S_\mathrm{qr}^\mathrm{d} S_\mathrm{qr}^\mathrm{d}]^{3/2}}{[S_\mathrm{lm} S_\mathrm{lm}]^{5/2} + [S_\mathrm{lm}^\mathrm{d} S_\mathrm{lm}^\mathrm{d}]^{5/4}}
    \qquad \qquad \text{with} \qquad \qquad  l^\mathrm{s} = C^\mathrm{w} \Delta \, , \label{equ:primal_wale}
\end{align}
where $C^\mathrm{w} = 0.35$ and $\Delta = (\Delta V)^{1/3}$ refer to the constant WALE coefficient as well as a local reference length aligned with the volume of a computational cell in the underlying Finite-Volume framework. The deviatoric strain rate tensor of the resolved scale reads
\begin{align}
    S_\mathrm{qr}^\mathrm{d} = \frac{1}{2}\left( \frac{\partial v_\mathrm{q} }{\partial x_\mathrm{m}} \frac{\partial v_\mathrm{m} }{\partial x_\mathrm{r}} + \frac{\partial v_\mathrm{r} }{\partial x_\mathrm{m}} \frac{\partial v_\mathrm{m} }{\partial x_\mathrm{q}}  \right) - \frac{1}{3} \delta_\mathrm{qr} \frac{\partial v_\mathrm{n} }{\partial x_\mathrm{m}} \frac{\partial v_\mathrm{m} }{\partial x_\mathrm{n}} \, .
\end{align}

\subsection{Detached Eddy Simulation}
\label{subsec:detached_eddy_simulation}

In line with the Improved Delayed Detached Eddy Simulation (IDDES) model of \cite{gritskevich2012development}, the turbulent viscosity is expressed as
\begin{align}
    \mu^\mathrm{t} &= \frac{\rho k}{\mathrm{max\left(\omega, f2 \frac{\sqrt{2 S_{ik} S_{ik}}}{a^{\mu}}\right)}} \, ,
\end{align}
where the required turbulent kinetic energy $k$ as well as its specific dissipation rate $\omega$ are obtained from the following transport equations
\begin{align}
	\frac{\partial \rho k}{\partial t} 
    + \frac{\partial}{\partial x_k} \left[ v_k \rho k 
    - \left( \mu + \mu^\mathrm{t} \, \sigma^\mathrm{k} \right) \frac{\partial k}{\partial x_k} \right] 
    - P^\mathrm{k} 
    + \rho \frac{k^{3/2}}{l^\mathrm{DES}} &= 0 \label{equ:tke_conservation} \\
    \frac{\partial \rho \omega}{\partial t} 
    + \frac{\partial}{\partial x_k} \left[ v_k \rho \omega
    - \left( \mu + \mu^\mathrm{t} \, \sigma^\mathrm{w} \right) \frac{\partial \omega}{\partial x_k} \right] 
    - 2 \sigma^\mathrm{\omega, 2} \frac{\rho}{\omega} \frac{\partial k}{\partial x_k} \frac{\partial \omega}{\partial x_k} \left( 1 - f1 \right)
    - \alpha P^\mathrm{k} \frac{\rho}{\mu^\mathrm{t}}
    + \beta \omega^2 \rho &= 0 \label{equ:omega_conservation}
\end{align}
where the TKE production is expressed as $P^\mathrm{k} = \mu^\mathrm{t} S_{ik} (\partial v_i / \partial x_k)$. Model coefficients
\begin{alignat}{3}
    \sigma^\mathrm{k} &=  \sigma^\mathrm{k,2}+\left( \sigma^\mathrm{k,1} - \sigma^\mathrm{k,2} \right) f1
    \qquad \qquad
    &&\sigma^\mathrm{\omega} &&= \sigma^\mathrm{\omega, 2}+\left( \sigma^\mathrm{\omega, 1} - \sigma^\mathrm{\omega, 2} \right) f1 \\
    \alpha &= \alpha^{\omega, 2} +\left( \alpha^{\omega, 1} - \alpha^{\omega, 2} \right) f1
    \qquad \qquad
    &&\beta &&= \beta^{\omega, 2} +\left( \beta^{\omega, 1} - \beta^{\omega, 2} \right) f1
\end{alignat}
as well as the denominator in the eddy viscosity are hyperbolically blended, viz.
\begin{align}
    f1 &= \mathrm{tanh} \left( \mathrm{min}\left(\mathrm{max}\left(
    \frac{ \sqrt{k}}{\beta^* \, \omega \, d},
    \frac{500 \mu}{\rho \, \omega \, d^2}
    \right),
    \frac{200 k \, \omega}{d^2 \, \frac{\partial k}{\partial x_k} \frac{\partial \omega}{\partial x_k}}
    \right)^4 \right) \\
    f2 &= \mathrm{tanh} \left( \mathrm{max}\left( 
    \frac{2 \sqrt{k}}{\beta^* \, \omega \, d},
    \frac{500 \mu}{\rho \, \omega \, d^2}
    \right)^2 \right) \, ,
\end{align}
where $d$ refers to the distance to the nearest wall. The following Shear Stress Transport (SST, cf. \cite{menter1994two}) model coefficients are employed
\begin{alignat}{7}
    \sigma^{k,1} &= 0.85 \qquad
    &&\sigma^{k,2} &&= 1.0 \qquad
    &&\sigma^{\omega,1} &&= 0.5 \qquad
    &&\sigma^{\omega,2} &&= 0.856 \\
    \alpha^{\omega,1} &= 5/9 \qquad
    &&\alpha^{\omega,2} &&= 0.44 \qquad
    &&\beta^{\omega,1} &&= 3/40 \qquad
    &&\beta^{\omega,2} &&= 0.0828 \\
    a^{\mu} &= 0.31 \qquad
    &&\beta^* &&= 0.09 \, .
\end{alignat}
The DES length scale determination follows \cite{gritskevich2012development}, i.e.,
\begin{align}
    l^\mathrm{DES} &= \tilde{f}^\mathrm{d} \, l^\mathrm{RANS} + (1 - \tilde{f}^\mathrm{d}) \, l^\mathrm{LES}
    \qquad \qquad \mathrm{with} \qquad \qquad
    l^\mathrm{RANS} = \frac{\sqrt{k}}{\omega \beta^*}
    \qquad \mathrm{and} \qquad
    l^\mathrm{LES} = C^\mathrm{DES} \Delta \, ,
\end{align}
where $\Delta = \mathrm{min}(C^\mathrm{d} \, \mathrm{max}(d, h), h)$ accounts for a local grid spacing that either refers to maximum edge length $h := h(\Delta V)$ per computational cell in the underlying Finite-Volume framework or a scaled ($C^\mathrm{d} = 0.15$) distance to the nearest wall. The LES's length scale coefficient is linear interpolated
\begin{align}
    C^\mathrm{DES} = C^\mathrm{DES, 1} f1 + C^\mathrm{DES, 2} (1 - f1) \, ,
\end{align}
and the blending between RANS and LES is performed via
\begin{align}
    \tilde{f}^\mathrm{d} = \mathrm{max}\left(\left(1 - f^\mathrm{d}\right),f^\mathrm{b}\right)
    \qquad \qquad \mathrm{with} \qquad \qquad
    f^\mathrm{d} = 1 - \mathrm{tanh}\left( \left(C^\mathrm{d, 1} \, r^\mathrm{d} \right)^{C^\mathrm{d, 2}}\right) \, ,
    f^\mathrm{b} = \mathrm{min}\left( 2 e^{-9 \alpha^\mathrm{fb}}, 1 \right) \, .
\end{align}
The required exponent reads
\begin{align}
    r^\mathrm{d} = \frac{\mu^\mathrm{t}}{\rho (\kappa d)^2 \sqrt{0.5} S \, W } 
\end{align}
and accounts for strain $S = \sqrt{2 S_{ik} S_{ik}}$ and vorticity $W = \sqrt{2 W_{ik} W_{ik}}$ rate magnitude, respectively, where $W_{ik} = 0.5 (\partial v_i / \partial x_k - \partial v_k / \partial x_i)$. The employed model constants are assigned to
\begin{align}
    C^\mathrm{DES, 1} = 0.78 \qquad \qquad
    C^\mathrm{DES, 2} = 0.61 \qquad \qquad
    C^\mathrm{d, 1} = 20 \qquad \qquad
    C^\mathrm{d, 2} = 3 \qquad \qquad
    \kappa = 0.41 \, .
\end{align}

\section{Instantaneous Results}

\subsection{Circular Cylinder}
\label{app:circular_cylinder}

\begin{figure}[!ht]
\centering
\iftoggle{tikzExternal}{
\input{./tikz/cylinder/cylinder_prope1_error.tikz}
}{
\includegraphics{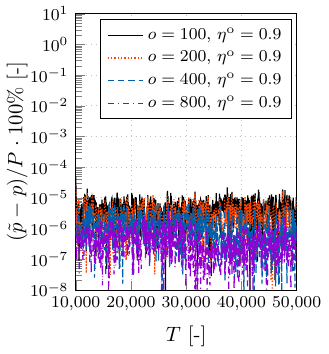}
\includegraphics{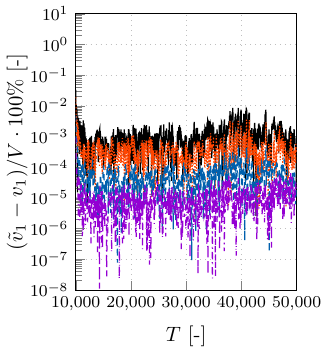}
\includegraphics{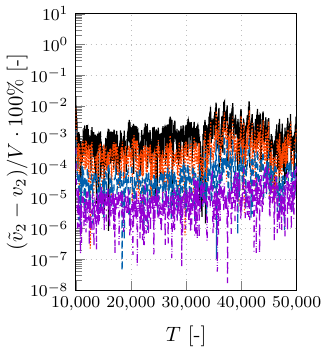}
}
\caption{Cylinder flow ($\mathrm{Re}_\mathrm{D} = 1.4 \cdot 10^5$): Instantaneous erros ($(\tilde{\varphi} - \varphi)/\varphi^\mathrm{ref} \cdot 100\%$) of reconstructed pressure (left), horizontal (center), and vertical (right) velocity for minimal construction ranks $o=[100, 200, 400, 800]$ with a retained energy of $\eta^\mathrm{o} = 0.9$ at probe 1 ($x_\mathrm{1}/D = -1.0$, $x_\mathrm{2}/D = 0.0$, $x_\mathrm{3}/D = 1.0$).}
\label{fig:cylinder_local_errors_probe_1}
\end{figure}
\begin{figure}[!ht]
\centering
\iftoggle{tikzExternal}{
\input{./tikz/cylinder/cylinder_prope3_error.tikz}
}{
\includegraphics{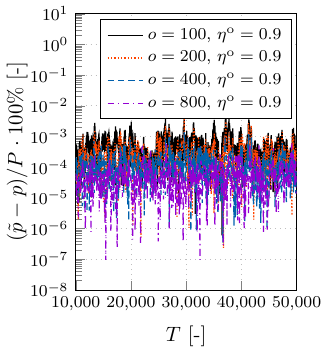}
\includegraphics{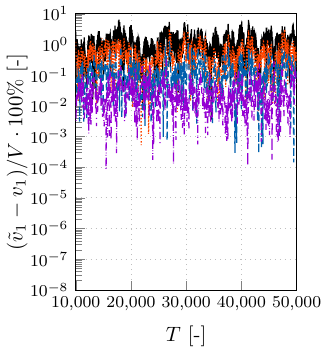}
\includegraphics{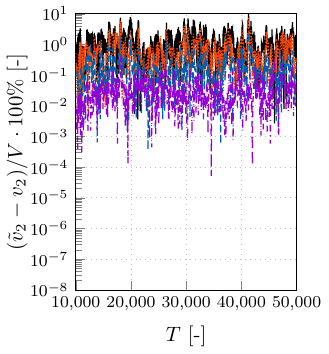}
}
\caption{Cylinder flow ($\mathrm{Re}_\mathrm{D} = 1.4 \cdot 10^5$): Instantaneous erros ($(\tilde{\varphi} - \varphi)/\varphi^\mathrm{ref} \cdot 100\%$) of reconstructed pressure (left), horizontal (center), and vertical (right) velocity for minimal construction ranks $o=[100, 200, 400, 800]$ with a retained energy of $\eta^\mathrm{o} = 0.9$ at probe 3 ($x_\mathrm{1}/D = 1.0$, $x_\mathrm{2}/D = -1.0$, $x_\mathrm{3}/D = 1.0$).}
\label{fig:cylinder_local_errors_probe_3}
\end{figure}
\begin{figure}[!ht]
\centering
\iftoggle{tikzExternal}{
\input{./tikz/cylinder/cylinder_prope4_error.tikz}
}{
\includegraphics{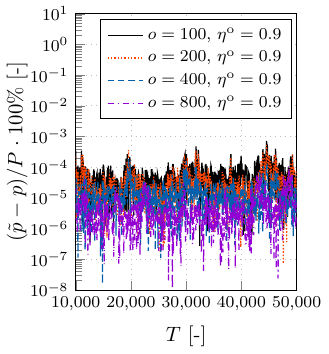}
\includegraphics{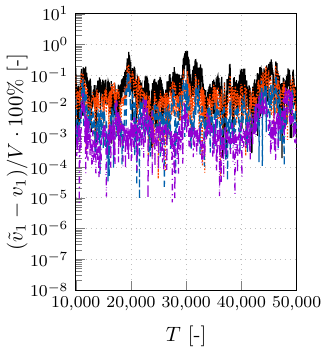}
\includegraphics{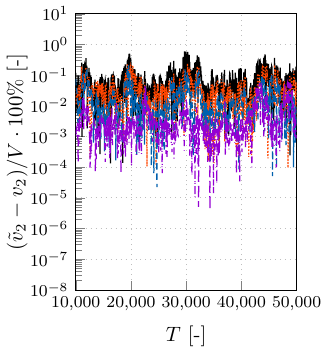}
}
\caption{Cylinder flow ($\mathrm{Re}_\mathrm{D} = 1.4 \cdot 10^5$): Instantaneous erros ($(\tilde{\varphi} - \varphi)/\varphi^\mathrm{ref} \cdot 100\%$) of reconstructed pressure (left), horizontal (center), and vertical (right) velocity for minimal construction ranks $o=[100, 200, 400, 800]$ with a retained energy of $\eta^\mathrm{o} = 0.9$ at probe 4 ($x_\mathrm{1}/D = 5.0$, $x_\mathrm{2}/D = 0.0$, $x_\mathrm{3}/D = 1.0$).}
\label{fig:cylinder_local_errors_probe_4}
\end{figure}

\newpage
\subsection{Feeder Ship}
\label{app:fleeder_ship}

\begin{figure}[!ht]
\centering
\iftoggle{tikzExternal}{
\input{./tikz/kcs/kcs_container_error_f1.tikz}
}{
\includegraphics{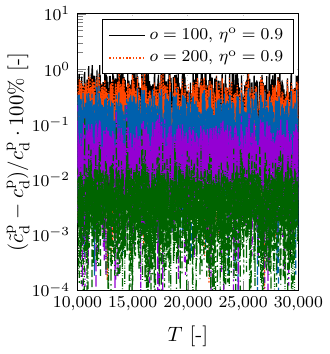}
\includegraphics{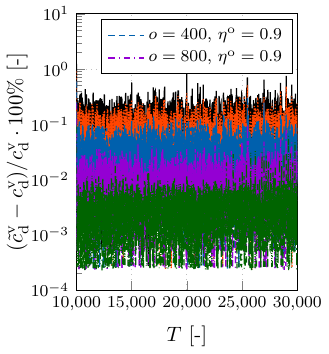}
\includegraphics{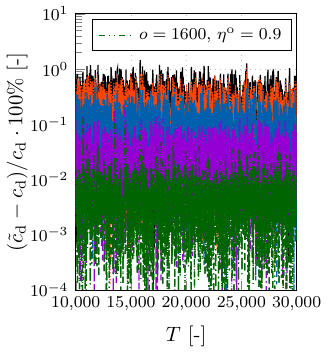}
}
\caption{Feeder ship flow ($\mathrm{Re}_\mathrm{L} = 5.0 \cdot 10^8$); Fifht container: Instantaneous relative error of pressure (left), viscous (center), and total (right) drag coefficient over the simulated time steps T for varying minimal tSVD ranks $o = [100, 200, 400, 800, 1600]$ for a retained energy level of $\eta^\mathrm{o} = 0.9$.}
\label{fig:kcs_container_error_f1}
\end{figure}
\begin{figure}[!ht]
\centering
\iftoggle{tikzExternal}{
\input{./tikz/kcs/kcs_deckshouse_error_f1.tikz}
}{
\includegraphics{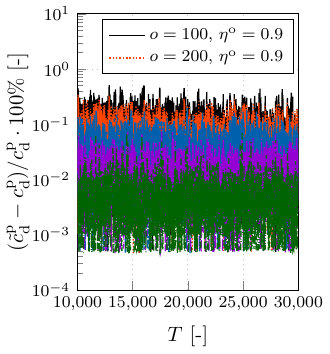}
\includegraphics{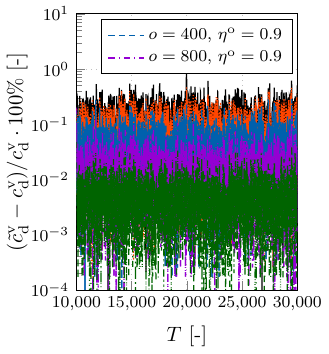}
\includegraphics{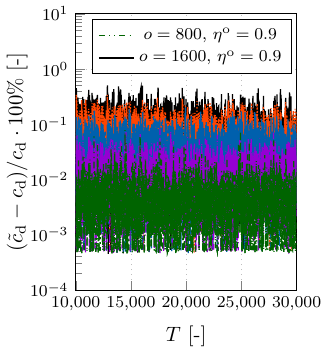}
}
\caption{Feeder ship flow ($\mathrm{Re}_\mathrm{L} = 5.0 \cdot 10^8$); Deckhouse: Instantaneous relative error of pressure (left), viscous (center), and total (right) drag coefficient over the simulated time steps T for varying minimal tSVD ranks $o = [100, 200, 400, 800, 1600]$ for a retained energy level of $\eta^\mathrm{o} = 0.9$.}
\label{fig:kcs_deckhouse_error_f1}
\end{figure}

\begin{figure}[!ht]
\centering
\iftoggle{tikzExternal}{
\input{./tikz/kcs/kcs_prope_star_1.tikz}
}{
\includegraphics{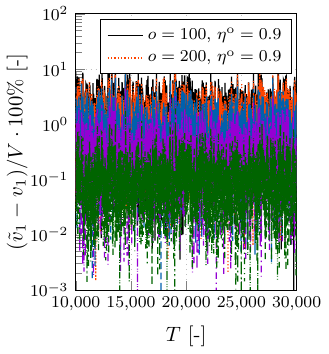}
\includegraphics{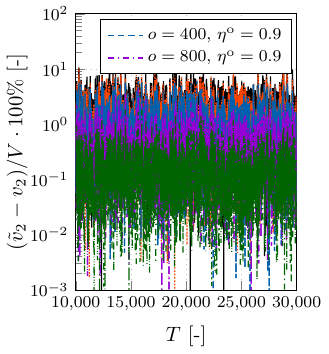}
\includegraphics{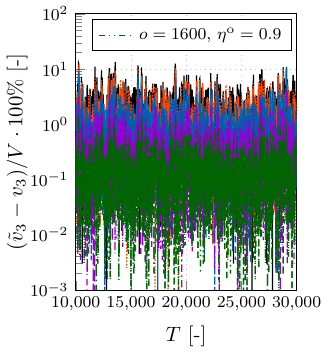}
}
\caption{Feeder ship flow ($\mathrm{Re}_\mathrm{L} = 5.0 \cdot 10^8$); Star side probe: Instantaneous error of longitudinal (left), lateral (center), and vertical (right) velocity over the varying minimal tSVD ranks $o = [100, 200, 400, 800, 1600]$ for different retained energy levels $\eta^\mathrm{o} \cdot 100\% = [10, 50, 90]$.}
\label{fig:kcs_prope_star_error_1}
\end{figure}
\begin{figure}[!ht]
\centering
\iftoggle{tikzExternal}{
\input{./tikz/kcs/kcs_prope_star_2.tikz}
}{
\includegraphics{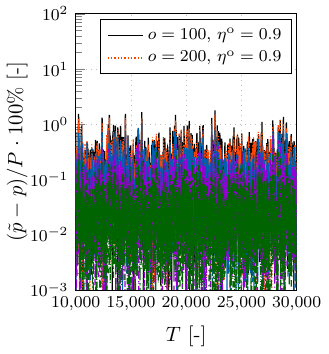}
\includegraphics{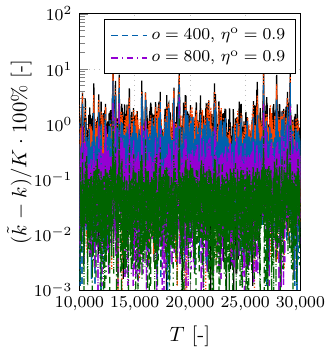}
\includegraphics{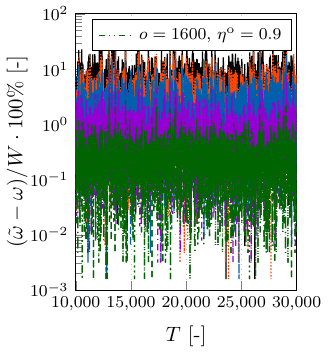}
}
\caption{Feeder ship flow ($\mathrm{Re}_\mathrm{L} = 5.0 \cdot 10^8$); Star side probe: Instantaneous error of pressure (left), TKE (center), and SDR (right) over the varying minimal tSVD ranks $o = [100, 200, 400, 800, 1600]$ for different retained energy levels $\eta^\mathrm{o} \cdot 100\% = [10, 50, 90]$.}
\label{fig:kcs_prope_star_error_2}
\end{figure}

\begin{figure}[!ht]
\centering
\iftoggle{tikzExternal}{
\input{./tikz/kcs/kcs_prope_port_1.tikz}
}{
\includegraphics{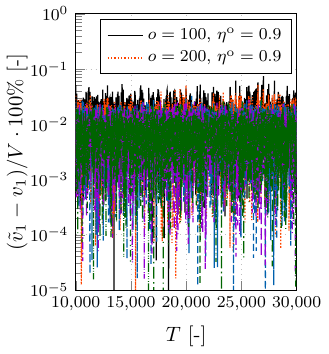}
\includegraphics{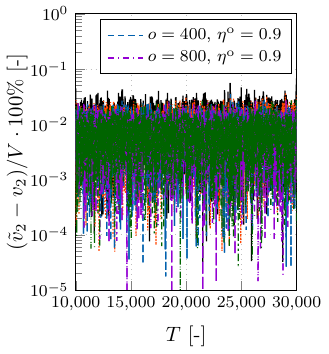}
\includegraphics{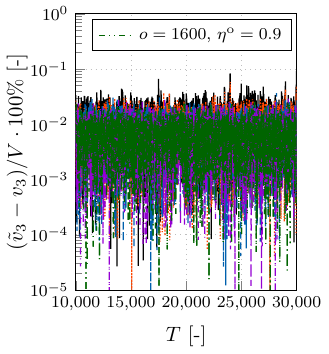}
}
\caption{Feeder ship flow ($\mathrm{Re}_\mathrm{L} = 5.0 \cdot 10^8$); Star port probe: Instantaneous error of longitudinal (left), lateral (center), and vertical (right) velocity over the varying minimal tSVD ranks $o = [100, 200, 400, 800, 1600]$ for different retained energy levels $\eta^\mathrm{o} \cdot 100\% = [10, 50, 90]$.}
\label{fig:kcs_prope_port_error_1}
\end{figure}
\begin{figure}[!ht]
\centering
\iftoggle{tikzExternal}{
\input{./tikz/kcs/kcs_prope_port_2.tikz}
}{
\includegraphics{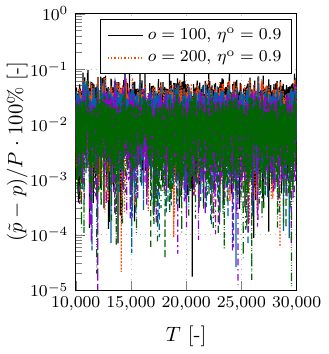}
\includegraphics{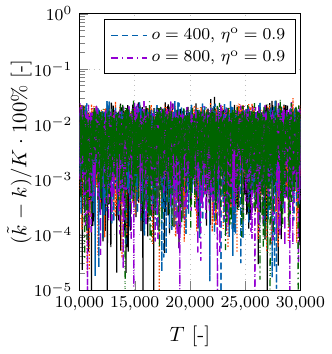}
\includegraphics{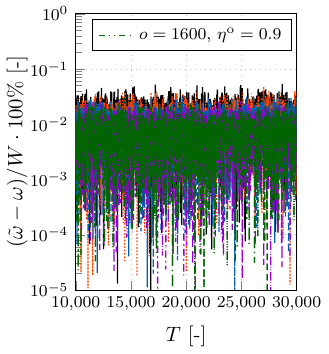}
}
\caption{Feeder ship flow ($\mathrm{Re}_\mathrm{L} = 5.0 \cdot 10^8$); Port side probe: Instantaneous error of pressure (left), TKE (center), and SDR (right) over the varying minimal tSVD ranks $o = [100, 200, 400, 800, 1600]$ for different retained energy levels $\eta^\mathrm{o} \cdot 100\% = [10, 50, 90]$.}
\label{fig:kcs_prope_port_error_2}
\end{figure}


\end{appendix}

\end{document}